\documentclass{tcibook}
\usepackage{fancyhea}
\usepackage{work}
\usepackage{amsmath,amssymb}

\usepackage{epsfig}  
\usepackage{graphicx}               % Standard graphics package  
\usepackage{url}
\usepackage{hyperref}
\newcommand{\nc}{\newcommand}  

\nc{\beq}{\begin{equation}}  
\nc{\eeq}{\end{equation}}  
\nc{\beqa}{\begin{eqnarray}}  
\nc{\eeqa}{\end{eqnarray}}  
\nc{\bea}{\begin{eqnarray}}  
\nc{\eea}{\end{eqnarray}}  
\nc{\ra}{\rightarrow}  
\nc{\lsim}{\begin{array}{c}\,\sim\vspace{-21pt}\\< \end{array}}  
\nc{\gsim}{\begin{array}{c}\sim\vspace{-21pt}\\> \end{array}}  
\nc{\slsh}{\slash\hspace*{-0.22cm}}

\def\to{\rightarrow}
\def\Re{{\cal R \mskip-4mu \lower.1ex \hbox{\it e}\,}}
\def\Im{{\cal I \mskip-5mu \lower.1ex \hbox{\it m}\,}}
\def\be{\begin{equation}}
\def\ee{\end{equation}}
\def\bea{\begin{eqnarray}}
\def\eea{\end{eqnarray}}
\def\bit{\begin{itemize}}
\def\eit{\end{itemize}}
\nc{\eref}[1]{(\ref{#1})}
\nc{\Eref}[1]{Eq.~(\ref{#1})}

\nc{\vev}[1]{ \left\langle {#1} \right\rangle }
\nc{\bra}[1]{ \langle {#1} | }
\nc{\ket}[1]{ | {#1} \rangle }
\nc{\fb}{\,{\rm fb}^{-1}}
\nc{\ev}{{\rm eV}}
\nc{\kev}{{\rm keV}}
\nc{\Mev}{{\rm MeV}}
\nc{\gev}{{\rm GeV}}
\nc{\tev}{{\rm TeV}}
\nc{\mev}{{\rm MeV}}

\def\ee{e^+e^-}

\def\msb{{\bar{\ssstyle M \kern -1pt S}}}

\begin{document}
\def\bibname{References}
\bibliographystyle{plain}

\raggedbottom
\pagenumbering{roman}
\parindent=0pt
\parskip=8pt
\setlength{\evensidemargin}{0pt}
\setlength{\oddsidemargin}{0pt}
\setlength{\marginparsep}{0.0in}
\setlength{\marginparwidth}{0.0in}
\marginparpush=0pt

% The content begins here 

\pagenumbering{arabic}
\renewcommand{\chapname}{chap:intro_}
\renewcommand{\chapterdir}{.}
\renewcommand{\arraystretch}{1.25}
\addtolength{\arraycolsep}{-3pt}

%%%%%%%%%% Contributor Defined Macros %%%%%%%%%%%%
%
%%%%% PSI n EDM

\newcommand{\mrm}{\mathrm}
\newcommand{\sub}[1]{\mathrm{\scriptscriptstyle{#1}}}

\newcommand{\lampHg}{\ensuremath{{}^{204}\mathrm{Hg}}}
\newcommand{\magHg}{\ensuremath{{}^{199}\mathrm{Hg}}}

\newcommand{\parafield}{\ensuremath{\uparrow\!\uparrow}}
\newcommand{\aparafield}{\ensuremath{\uparrow\!\downarrow}}

\newcommand{\nedm}{\ensuremath{d_{\sub{n}}}}
\newcommand{\hgedm}{\ensuremath{d_{\sub{Hg}}}}
\newcommand{\ecm}{\ensuremath{e\!\cdot\!\mathrm{cm}}}
\newcommand{\ntwoedm}{\ensuremath{\mathrm{n^2EDM}}}
\newcommand{\fTHz}{\ensuremath{\mathrm{fT/\sqrt{Hz}}}}

\newcommand{\tsups}[1]{\textsuperscript{#1}}
\newcommand{\trinat}{{\scshape Trinat}}
\newcommand{\triumf}{{\scshape Triumf}}
\newcommand{\trex}{T{\small REX}}
\newcommand{\zerotozero}{\mbox{$0^+\!\!\rightarrow 0^+\ $}}
\newcommand{\tamutrap}[0]{{\scshape Tamutrap}}

%%    TEMPLATE for contributions to the Proceedings of the
%%       Workshop on Fundamental Physics at the Intensity Frontier
%%    
%%     
%%
%%
%%     Questions?  Send email to : hewett@slac.stanford.edu  %%   
%\def\Title#1{\begin{center} {\large {\bf #1}} \end{center} 

%\hfill{ACFI-T13-04}

\chapter{Fundamental Symmetry Tests with Nucleons, Nuclei, and Atoms: A Snowmass Report}
\label{chap:chapx}

\begin{center}
{Conveners: Krishna~Kumar,$^{1}$ Zheng-Tian~Lu,$^{2,3}$ Michael~J.~Ramsey-Musolf$^{1,4,5}$}
\smallskip
$^{1}$Department~of~Physics, University~of~Massachusetts, Amherst, MA, USA\\
$^{2}$Physics~Division, Argonne~National~Laboratory, Argonne, IL, USA\\
$^{3}$Department~of~Physics, The~University~of~Chicago, Chicago, IL, USA\\
$^{4}$Amherst~Center~for~Fundamental~Interactions, University~of~Massachusetts, Amherst, MA, USA\\
$^{5}$Kellogg~Radiation~Laboratory, California~Institute~of~Technology, Pasadena, CA, USA\\
\end{center}

\vskip 0.25in
{\bf Abstract:} Present and prospective fundamental symmetry tests with nucleons, nuclei and atoms are probing for possible new physics at the TeV scale and beyond.
These ongoing and proposed table-top as well as accelerator-based experiments are thus a vital component of the Intensity Frontier. At the same time, these tests provide increasingly sophisticated probes of long-distance strong interactions that are responsible for the structure of nucleons and nuclei. In this community report, some of the most compelling opportunities with nucleons, nuclei and atoms are summarized, drawing largely on input received from the nuclear and atomic physics communities. In particular, this report includes many contributions submitted to two recent Intensity Frontier Workshops.

UMass preprint: ACFI-T13-04

\vskip 0.25in

\tableofcontents
\newpage                                                        
%%%%%%%%%%%%%%%%%%%%%%%%%%%%%%%%%%%
\begin{center}\begin{boldmath}
%\Large{list of authors}
\end{boldmath}\end{center}

\section{Overview}\label{sec:NNAintro}
%\section{Introduction}\label{sec:NNAintro}

Despite the success of the Standard Model in explaining so many subatomic physics phenomena, we recognize that
the model is incomplete and must eventually give way to a more fundamental description of nature.  We have discovered
massive neutrinos and associated flavor violation, which require the introduction of new mass terms in the Standard Model.
We have an excess of baryons over antibaryons in our universe, indicating baryon-number-violating interactions
and likely new sources of CP violation.  We know from the inventory of matter and energy in our universe that the
portion associated with Standard Model physics is only about 5\% of the total.  The rest remains unidentified and quite
mysterious.

These \lq\lq big questions" -- the origin of matter, the nature of neutrino mass, the identification of dark matter and dark
energy - have driven the two major thrusts of subatomic physics.  One is the effort to probe ever shorter distance scales
by advancing the energy frontier.  Today that frontier in represented by CERN's LHC.   The alternative is to seek 
signals of the new physics in subtle violations of symmetry in our low-energy world -- ultraweak interactions that
might mediate lepton- or baryon-number violation, generate electric dipole moments, or lead to unexpected flavor
physics.  This second approach is the theme of this chapter:  ultrasensitive techniques in atomic, nuclear, and
particle physics that might reveal the nature of our next standard model.  Like particle astrophysics
and cosmology, the third leg of our
efforts to find new physics, this second approach uses our world as a laboratory, and depends on precision to
identify the new physics.

This field has a long and successful history. Tests of fundamental symmetries in experiments involving nucleons and nuclei have played an essential role in developing and testing the Standard Model. In a landmark experiment in 1956, the observation of parity-violation in the radioactive decay of $^{60}$Co, shortly preceding the observation of parity-violation in pion decay, provided the first experimental evidence that the weak interaction does not respect this symmetry, ultimately leading to the Standard Model description of charged weak currents as being purely left-handed. Similarly, the measurements of the parity-violating asymmetry in polarized deep-inelastic electron-deuteron scattering in the 1970's  singled out the Standard Model structure for the neutral weak current from among competing alternatives, well in advance of the discovery of the electroweak gauge bosons at CERN. And the non-observation of a permanent electric dipole moment (EDM) of the neutron and $^{199}$Hg atoms has placed stringent bounds on the possibility of combined parity and time-reversal violation (or CP-violation) in the strong interaction, motivating the idea of the spontaneously-broken Peccei-Quinn symmetry and the associated axion that remains a viable candidate for the cosmic dark matter. 

Present and prospective fundamental symmetry tests with nucleons, nuclei and atoms are now poised to probe for possible new physics at the Terascale and beyond, making them a vital component of the Intensity Frontier. At the same time, these tests provide increasingly sophisticated probes of long-distance strong interactions that are responsible for the structure of nucleons and nuclei. The potential for both discovery and insight has motivated the nuclear physics community to identify studies of fundamental symmetries and neutron properties as one of the top-four priorities for the field in the 2007 Nuclear Science Advisory Long Range Plan \cite{lrp2007}, perhaps anticipating the present broader interest in the Intensity Frontier that underlies this document. Below, some of the most compelling opportunities with nucleons, nuclei and atoms are summarized, drawing largely on input received from the nuclear and atomic physics communities.

Before proceeding, we note that the bulk of the discussion that follows draws heavily on input from the community provided at the Intensity Frontier Workshop held in Rockville, MD (November 30-December 2, 2011), and a second workshop held at Argonne National Laboratory (April 24-27, 2013). In particular, a series of two-page \lq\lq mini white papers" written by researchers in the field have been adapted into various sections of this document. The contributors are J. Behr, D.A. Bryman, R. Carlini, T.E. Chupp, A. Deshpande, J. Fajans, P. Fierlinger, B.W. Filippone, A. Garcia, S. Gardner, H.A. Gould, V.P. Gudkov, J.C. Hardy, R.J. Holt, Y. Kamyshkov, K. Kirch, A.D. Krisch, A.S. Kronfeld, C.Y. Liu, W. Lorenzon, J.W. Martin, D. Melconian, O. Naviliat-Cuncic, J.S. Nico, L.A. Orozco, Y.K. Semertzidis, D. Sivers, P.A. Souder, E. Widmann, A. Young.

We have made no attempt to fill in all the gaps resulting from this process, and no editorial judgment is implied on the relative priority of what is discussed below or omitted. We also emphasize that the discussion is largely focused on experimental opportunities, though the vital role of theoretical work in guiding and interpreting the experimental program should not be overlooked. Throughout, we refer briefly to some of the open theoretical questions. For a more detailed discussion, we refer the reader to a series of nine review articles has that appeared in a dedicated volume of Progress in Particle and Nuclear Physics\cite{Cirigliano:2013lpa,Engel:2013lsa,Cirigliano:2013xha,Erler:2013xha,deGouvea:2013zba,Balantekin:2013tqa,Balantekin:2013gqa,Gardner:2013ama,Haxton:2013aca}, following from a meeting on Beyond the Standard Model in Nuclear Physics  held at the University of Wisconsin-Madison in October, 2011 and sponsored jointly by the DOE and NSF nuclear physics program offices.

\section{Electric Dipole Moments}
\label{sec:EDM}

\begin{figure}[htbp]
\centering
\vspace*{3mm}
\includegraphics[width=0.4\textwidth]{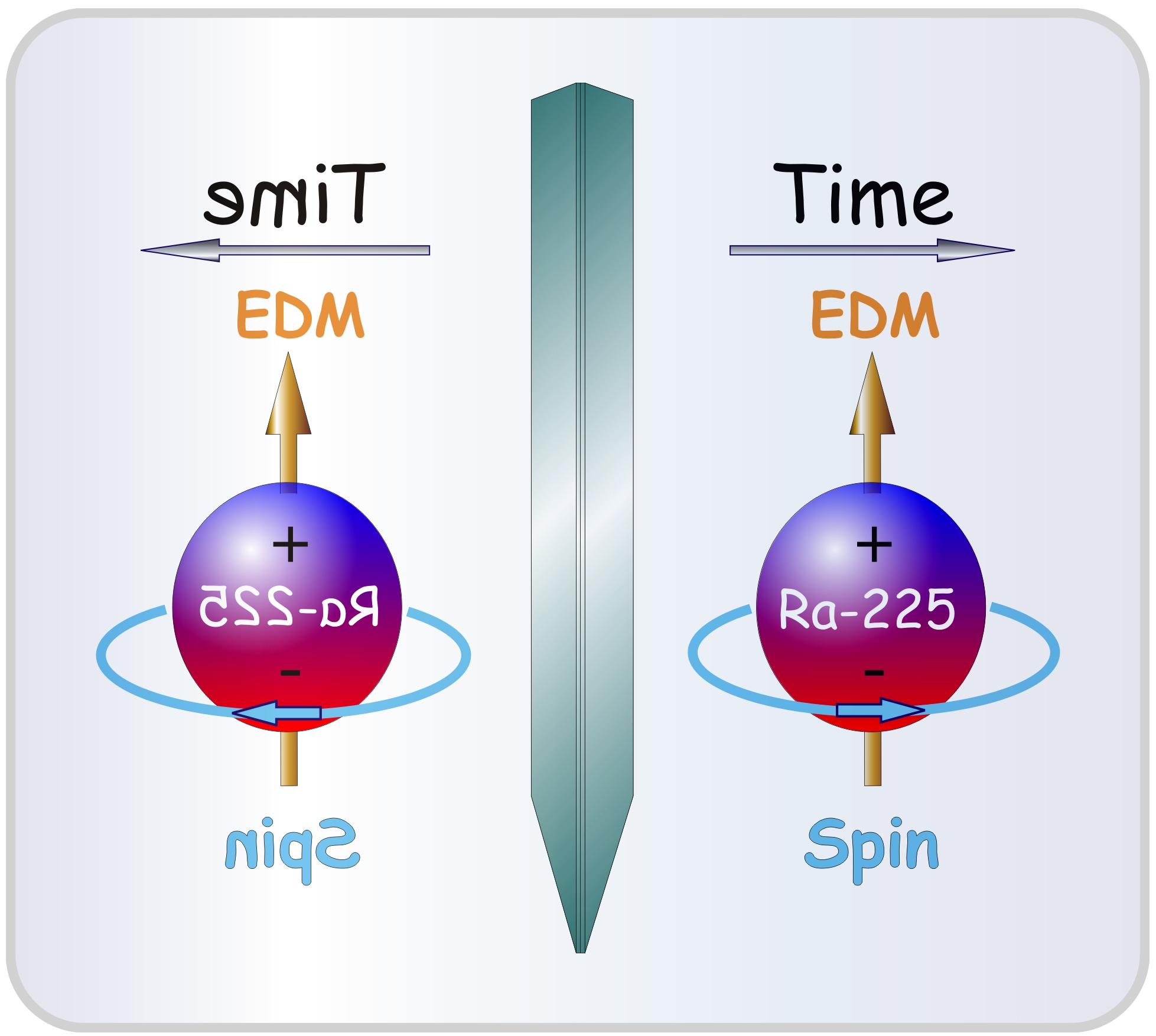}
\caption{The interaction of an elementary particle EDM with an electric field  violates time-reversal symmetry. When time is reversed, the spin is reversed,  and so is the elementary particle EDM. The EDM of a classical system that corresponds to a spatial charge separation (denoted by \lq\lq EDM" in the figure)  does not reverse. Thus, the classical EDM interaction in a time-reversed world looks the same as in the original world, while the elementary particle EDM interaction does not.}
\label{EDM1}
\end{figure}

In an heuristic but na\"I've picture, the permanent electric dipole moment (EDM) of a particle arises from the spatial separation of opposite charges along the axis of the particle's angular momentum. The existence of an EDM would be a direct signature of the violation of both parity (P) and time-reversal symmetry (T) (Fig. \ref{EDM1}). It would also probe physics of CP violation (C stands for charge conjugation) which necessarily accompanies T violation under the assumption of the CPT theorem. EDM measurements conducted in many laboratories around the world employing a variety of techniques have made tremendous progress, and all have so far obtained results consistent with zero EDM. For example, in the past six decades, the search sensitivity of the neutron EDM has improved by six orders of magnitude to reach the current upper limit of $2.9 \times 10^{-26} \ecm$ \cite{Bak06}.

CP violation in flavor-changing decays of K- and B-mesons have been observed. The results can be explained, and indeed in many cases predicted, by the CKM mechanism within the framework of the Standard Model. All of the observed CP violation phenomena originate from a single complex phase in the CKM matrix that governs the mixing of quark flavors. While this elegant solution has been well established following a string of precise measurements, additional sources of CP violation are generally anticipated in extensions of the Standard Model. For example, in SUSY, the supersymmetric partners of quarks would naturally allow additional complex phases in the expanded mixing matrix and induce new CP-violating phenomena. Moreover, a CP-violating term is known to be allowed in the general form of the QCD Lagrangian, and would have CP-violating consequences specifically in the strong interaction. Additional CP-violating mechanisms are also called for by the observation that the baryon-to-photon ratio in the Universe is as much as nine orders of magnitude higher than the level that can be accommodated by the Standard Model. A much more significant matter-antimatter asymmetry is likely to have been present in the early Universe and provided the favorable condition for the survival of matter that we observe today.

EDMs are  sensitive probes for new CP-violating mechanisms, and the collection of EDM searches is generally considered to be one of the most promising paths towards new physics beyond the Standard Model. The CKM mechanism in the Standard Model can only generate EDMs at the three- and four- loop level, leading to values many orders of magnitude lower than the current experimental limits. For example, the Standard Model value for the neutron EDM is expected to be approximately $10^{-31}$ \ecm, or five orders of magnitude below the current upper limit. Any non-zero EDM observed in the foreseeable future would have to require either CP-violation in the strong interaction or physics beyond the Standard Model. Perhaps unsurprisingly, extensions of the Standard Model generally allow a range of EDM values that are within the reach of the on-going experiments. The scientific importance and discovery potential of EDM searches are strongly endorsed by the communities of both particle physics and nuclear physics. The Nuclear Science Advisory Committee (NSAC) proclaimed in the 2007 Long Range Plan~\cite{lrp2007} that ``a nonzero EDM would constitute a truly revolutionary discovery.'' The negative findings so far are also valuable. As was pointed out in the 2006 P5 report~\cite{p5}, \emph{The Particle Physics Roadmap}, ``the non-observation of EDMs to-date, thus provides tight restrictions to building theories beyond the Standard Model.'' Specifically, the upper limits on EDM provide insight on the scale of the next energy frontier.

The most sensitive EDM searches have so far been conducted on the neutron, nuclei ($^{199}$Hg), and the electron (Table 1). On the one hand, experiments in these three categories all compete for the prize of being the first to observe a non-zero EDM. On the other hand, they complement each other as each category is most sensitive to different sources of CP violation. For example, the neutron is more sensitive to the EDMs of its constituent quarks; heavy nuclei are more sensitive to the quark chromo-EDM and other CP-violation mechanisms in the nuclear force. The recently proposed storage ring EDM experiments of the proton and deuteron aim to probe combinations of CP-violating contributions that differ from the neutron EDM.  In the future, if a non-zero EDM is discovered in one particular system, it would still be necessary to measure EDMs in other categories to help resolve the underlying CP-violation mechanisms. 

At the intensity frontier, a new generation of sources for cold neutrons and ultracold neutrons (UCNs) are becoming available. Their higher output in neutron flux will enable searches for the neutron EDM projected at a sensitivity level of 10$^{-28}$ \ecm, or two orders of magnitude below the current best limit. A survey of neutron EDM experiments are presented in this section. Also at the intensity frontier, future isotope production facilities such as FRIB after upgrade or Project X will produce a prolific amount of selected isotopes that possess enhanced sensitivities to the EDMs of the nuclei or the electron. Included in this section are the cases for the radium and radon isotopes.

\begin{table}
\centering
\caption{Upper limits on EDMs in three different categories. Note that the limit on the electron EDM from the ThO molecule assumes that only the EDM of the electron would contribute to a signal. In a  model independent analysis, the ThO limit constrains a linear combination of the electron EDM and a CP-violating semileptonic interaction (see, {\em e.g.}, Ref.~\cite{Engel:2013lsa}).}
\label{table1}
\begin{tabular}{|l|c|c|c|}
\hline
 Category & EDM Limit (\ecm)  & Experiment &  Standard Model value (\ecm)   \\
\hline
 Electron             & $8.7\times10^{-29}$    &   ThO molecules in a beam \cite{ACME:2013}  &   10$^{-38}$       \\
 Neutron             & $2.9\times10^{-26}$    &   Ultracold neutrons in a bottle \cite{Bak06}   &   10$^{-31}$       \\
 Nucleus              & $3.1\times10^{-29}$    &  $^{199}$Hg atoms in a vapor cell  \cite{Griffith:2009zz}   &   10$^{-33}$      \\
\hline
\end{tabular}
\end{table}

\subsection{PSI Neutron EDM}
\label{sec:PSIEDM}

A nEDM experiment is being developed in steps at the Paul Scherrer Institute (PSI)~\cite{Alt09a}. The collaboration is 
pursuing a considerable technical R\&D effort
but also exploiting the complementary physics potential of the nEDM
apparatus with respect to exotic interactions~\cite{Ban07}.
The experiment is located at the new source for ultracold neutrons (UCN) at
PSI~\cite{lauss,blau}. This source uses neutron production via
proton induced spallation on lead, moderation in heavy water and
solid deuterium, and downscattering to UCN. Through an intermediate
storage volume UCN can be distributed to three experimental beam ports. 
The performance of the source is continuously being optimized.
Besides nEDM, the UCN source can also serve other experiments.

The collaboration is presently using the original but upgraded
Sussex-RAL-ILL spectrometer~\cite{Bak06}. In its configuration at
the PSI UCN source, it is estimated to yield a factor of 25 higher
statistics as compared to the earlier ILL setup. This increased statistical
sensitivity needs to be accompanied by a comparable reduction of systematic uncertainties, particularly on the control and measurement of magnetic fields. The near-term goal is to accumulate enough data
to reach a sensitivity of $\sigma(\nedm)=2.6\times10^{-27}\,\ecm$, which
corresponds to an upper limit of $d_{\sub{n}}<
5\times10^{-27}\,\ecm$~(95\,\% C.L.) in case of a null result.

The next-generation neutron EDM
experiment at PSI, named n2EDM, is being designed and will be constructed and
offline tested in parallel to operating nEDM.  It will be operated at
room-temperature and in vacuum, aiming at a sensitivity of
$d_{\sub{n}}< 5\times10^{-28}\,\ecm$~\cite{Alt09a} (95\% C.L. limit
in case of no signal). The setup is built around two stacked neutron
precession chambers for simultaneous measurements of both E-field
orientations. Precise control and measurement of the magnetic
environment online inside the apparatus is possible via laser
read-out Hg co-magnetometers, multiple Cs magnetometers as
gradiometers surrounding the neutron precession chamber, and
additional $^3$He magnetometer cells both above and below the neutron
chambers. At present the setup area for the n2EDM apparatus is being prepared.

\subsection{ILL Neutron EDM}
\label{sec:ILLnEDM}

\paragraph{\bf CryoEDM:} The CryoEDM experiment at the Instut Laue-Langevin (ILL) in Grenoble, France, uses resonant downscattering of $9\AA$ neutrons in a bath of superfluid $^4$He as a source of UCN.  The UCNs are transported to magnetically shielded storage cells where, as in the previous generation of this experiment carried out at room-temperature, the Ramsey technique of separated oscillatory fields is used to measure the precession frequency of the neutron in parallel and antiparallel electric and magnetic fields.  There are two Ramsey chambers: one has no electric field applied, and serves as a control.  Magnetic-field fluctuations are monitored with SQUIDs.  The neutrons are counted using detectors situated within the liquid helium~\cite{Bak03}.  The experiment is in its commissioning stage.  Once it reaches the sensitivity of the room-temperature experiment~\cite{Bak06}, it will be moved to a new beamline, where upgrades to various components of the apparatus should lead to an improvement of about an order of magnitude in sensitivity.

\paragraph{\bf PNPI/ILL nEDM:} Also at ILL, a PNPI/ILL experiment~\cite{Ser09} to measure nEDM is currently being prepared at the UCN facility PF2. To enable an improvement of sensitivity, one of the PF2's beam positions has been equipped with new components for UCN transport, polarization and beam characterization, comprised of a superconducting solenoid-polarizer with a 4 Tesla magnetic field, a neutron guide system with a 136 mm diameter prepared in replica technology, and a novel beam chopper for time-of-flight analysis. The whole EDM apparatus is set up on a non-magnetic platform. A higher density of polarized UCNs at the experimental position, at approximately 5 $cm^{-3}$, shall lead to an EDM measurement with a counting statistical accuracy of  $1.5\times10^{-26}$ \ecm~during 200 days of operation at PF2.

\subsection{SNS Neutron EDM}
\label{sec:USnEDM}

The goal of the SNS nEDM experiment, to be carried out at the 
Spallation Neutron Source (SNS), is to achieve a sensitivity $< 3 \times
10^{-28}$ \ecm. A value (or limit) for the neutron EDM will be extracted from the difference
between neutron spin precession frequencies for parallel and anti-parallel
magnetic ($\sim 30$ mGauss) and electric ($\sim 70$ kV/cm) fields.
This experiment uses a novel 
polarized $^3$He co-magnetometer and will 
detect the neutron precession via the spin-dependent neutron capture on 
$^3$He~\cite{ref3}. The capture reaction produces energetic proton and triton, which
ionize liquid helium and generate scintillation light that can be detected.
Since the EDM of $^3$He is strongly suppressed by electron screening in the
atom it can be used as a sensitive magnetic field monitor. 
High densities of trapped UCNs are produced via
phonon production in superfluid $^4$He which can also support large 
electric fields. This technique allows for a number of independent 
checks on systematics including: 1) Studies of the temperature dependence of false EDM signals in the $^3$He; 2) Measurement of the $^3$He precession frequency using SQUIDs; 3) Cancellation of magnetic field fluctuations by matching the 
effective gyromagnetic ratios of neutrons and $^3$He with the ``spin dressing'' technique~\cite{ref3}.

The collaboration is continuing to address critical R\&D developments in preparation for construction of a full experiment. Key issues being addressed include: 1) Maximum electric field strength for large-scale electrodes made of
appropriate materials in superfluid helium below a temperature of 1 K. 2) Magnetic field uniformity for a large-scale magnetic coil and a superconducting Pb magnetic shield. 3) Development of coated measurement cells that preserve both neutron and 
$^3$He polarization along with neutron storage time. 4) Understanding of polarized $^3$He injection and transport in the superfluid. 5) Estimation of the detected light signal from the scintillation in superfluid helium.

The experiment will be installed at the FNPB (Fundamental Neutron Physics Beamline) at the SNS and construction is likely to take at least five years, followed by hardware commissioning and data taking. Thus first results could be anticipated by the end of the decade.

\subsection{TRIUMF Neutron EDM}
\label{sec:TRIUMFnEDM}

The basic design of the experiment calls for a room-temperature EDM
experiment to be connected to a cryogenic UCN source \cite{masuda}.  
The source will be operated at the Research
Center for Nuclear Physics (RCNP, Osaka) and then moved to TRIUMF.  The goal is to achieve $>$ 5000 UCN/cm$^3$ in an nEDM measurement cell.
A prototype nEDM apparatus has been characterized in beam tests
at RCNP Osaka.  Using this apparatus the collaboration has already demonstrated long
UCN storage lifetimes, polarization lifetimes, and transverse spin
relaxation times.

The EDM apparatus has a few unique features: A spherical
coil within a cylindrical magnetic shield is used to generate the DC magnetic
field;  a $^{129}$Xe comagnetometer is used to address false EDMs due to a geometric phase
effect; and due to the expected higher UCN density, 
the measurement cell size is designed to be considerably smaller than the previous ILL
apparatus.  While having a negative impact on statistics, the reduced
cell size limits systematic effects, particularly from the geometric phase effect.

The long-term goal, to be reached in 2018 and beyond, is $d_n<1\times
10^{-28}$~\ecm. Experiments on the neutron lifetime and
on neutron interferometry are also considered as candidates for the
long-term physics program.

\subsection{Munich Neutron EDM}
\label{sec:MunichnEDM}

At the new UCN source of FRM-II in Garching, Germany, a next-generation neutron EDM experiment aims to achieve a statistical limit of $d_n < 5\cdot 10^{-28}$~\ecm~at $3 \sigma$ and a corresponding control of systematic effects of $\sigma_{d,syst} < 2\cdot 10^{-28}$~\ecm ($1 \sigma$). 
The source of UCN is placed in a tangential beam tube inside the reactor with a thermal neutron flux of $10^{14}$~s$^{-1}$. 
Solid deuterium is used as a super-thermal converter for the production of UCN~\cite{Frei07}.
A beamline made from specially prepared replica foil tubes with a relative transmission of $> 0.99$ per meter guides the UCNs to the nEDM spectrometer, which is placed outside the reactor building in a new experiment hall at 27 m distance from the solid deuterium source. 
Taking into account production, volumes and losses of all components and the EDM chambers, the projected polarized UCN density is $>$ 3000 cm$^{-3}$ in the EDM experiment.

This experiment is based on UCN stored in two vertically aligned cylindrical vessels at room temperature and a vertical magnetic field $B_0$.  In between the cells a high voltage electrode is placed to enable measurements with an electric field parallel and anti-parallel to $B_0$ simultaneously.
For EDM measurements, Ramsey's method of separated oscillatory fields is applied to these trapped UCN.
With a precession time of $T = 250$~s,  an electric field $E = 18$~kV$/$cm, the statistical sensitivity goal can be achieved in 200 days.
In addition, a co-magnetometer based on polarized $^{199}$Hg vapor with a laser based optical system is placed in these cells~\cite{Griffith:2009zz}.
In additoin, external magnetometers are used to measure the field distribution online.
Buffer gases can be added to all magnetometers to investigate various systematic effects and to eventually increase the high voltage behavior.

\subsection{Proton Storage Ring EDM}
\label{sec:pEDM}

The storage ring EDM collaboration has submitted  a proposal to DOE for a proton EDM experiment sensitive
to $10^{-29} \ecm$~\cite{edmweb}. This experiment can be done at Brookhaven National Laboratory (BNL) or another facility 
that can provide highly polarized protons with an intensity of more than $10^{10}$ particles per cycle of 15 minutes. 
The method utilizes polarized protons at the so-called ``magic'' 
momentum of 0.7 GeV/{\it c} in an all-electric storage ring with a radius of $\sim 4$0 m.  At this momentum, the proton spin and 
momentum vectors precess at the same rate in any transverse electric field.  When the spin is kept along the momentum
direction, the radial electric field acts on the EDM vector causing the proton spin to precess vertically. The vertical component 
of the proton spin builds up for the duration of the storage time, which is limited to $10^{3}$ s
by the estimated horizontal spin coherence time (hSCT) of the beam within the admittance of the ring.

The strength of the storage ring EDM method comes from the fact that a large number of highly polarized particles can be stored
for a long time, a large hSCT can be achieved and the transverse spin components can be probed as a function of time with a high
sensitivity polarimeter.  The polarimeter uses elastic nuclear scattering off a solid carbon target placed in a 
straight section of the ring serving as the limiting aperture. The collaboration has over the years developed the method and 
improved their understanding and confidence on it.

\subsection{Mercury-199 Atomic EDM}
\label{sec:HgEDM}
The mercury atom provides a rich hunting ground for sources of CP violation. An EDM in $^{199}$Hg could be generated by EDMs of the neutrons, protons, or electrons, by chromo-edms of the quarks, by CP-odd electron-nucleon couplings, or by $\theta_{QCD}$, the CP-odd term in the strong interaction Lagrangian. The current upper limit on the Hg EDM~\cite{Griffith:2009zz},  $d(\mathrm{^{199}Hg})<3.1\times10^{-29}\ecm$, places the tightest of all limits on chromo-edms, the proton edm, and CP odd electron-nucleon couplings.

The statistical sensitivity of the current upper limit on $d(^{199}Hg)$ was limited by two noise sources: light shift noise and magnetic Johnson noise. The light shift noise was due to a combination of residual circular polarization of the probe light and a small projection of the probe light axis along the main magnetic field axis. This noise was subsequently reduced by a factor ten by better alignment of the probe light axis and will be further reduced by letting the atoms precess in the dark. The next data runs will be taken with the probe light on only at the start and end of the precession period. The magnetic Johnson noise was generated by thermally excited currents in the aluminum cylinder that held the windings of the main magnetic field coil. The aluminum coil form has been replaced by an insulating coil form, leaving magnetic field noise from the innermost magnetic shield as the dominant remaining noise source.  If the dominant noise source in the next data runs is indeed noise from the magnetic shield, then a factor of ten improvement in statistical sensitivity can be achieved with the existing Hg EDM apparatus.

An increase in statistical sensitivity requires a corresponding increase in the control of systematic errors. The dominant systematic error has been imperfect knowledge about the magnetic fields produced by leakage currents across the Hg vapor cells when high voltage is applied across the cells. Recently, it was found that most of the leakage current flows along electric field lines in the dry nitrogen gas exterior to the cells; these gas currents can be amplified and have been shown to not produce measurable systematic errors. Roughly 10\% of the total current flows along the cell walls and will be a source for concern. However, by maintaining these cell wall leakage currents below 0.01 pA, as has been achieved in earlier EDM measurements, a ten-fold improvement in the leakage current systematic error can be achieved.

In summary, unless unforeseen problems emerge, the existing Hg EDM apparatus can provide a ten-fold increase in sensitivity to an Hg atom EDM. This would still be roughly a factor of ten larger than the shot noise limit of the current apparatus. If warranted, a new apparatus could be developed to go further. A larger diameter and thicker walled innermost magnetic shield would reduce the magnetic field noise and additional magnetometers could be installed to provide further information about the field stability. Redesigned vapor cells could reduce the leakage currents and better direct their paths (eg rectangular cells with a reduced electric field gap). An additional factor of five increase in sensitivity would be feasible.

\subsection{Radon-221,223 Atomic EDM}
\label{sec:RnEDM}

In a heavy atom of a rare isotope, for which the nucleus has octupole strength or permanent deformation, the dipole charge distribution in the nucleus, characterized by the Schiff moment, may be significantly enhanced compared to $^{199}$Hg. 
This enhancement is due to the parity-odd moment arising from quadrupole-octupole interference, and the enhanced E1 polarizability effected by closely spaced levels of the same $J$ and opposite parity.   The strongest octupole correlations occur near $Z=88$ and $N=134$, and isotopes $^{221/223}$Rn and $^{225}$Ra are promising for both practical experimental reasons and as candidates for octupole-enhanced Schiff moments. Enhancements of the nuclear Schiff moment by a factor of 100 or more compared to $^{199}$Hg have been predicted by models using Skyrme-Hartree-Fock for $^{225}$Ra~\cite{rf:deJesus2005} and Woods-Saxon and Nilsson potentials in the case of $^{223}$Rn~\cite{rf:Spevak1997}. However, the uncertainties on the size of enhancements are quite large, in part due to uncertainty in the $^{199}$Hg Schiff moment, and, in the case of $^{221/223}$Rn isotopes, the absence of nuclear structure data. A recent measurement of the E2 and E3 intrinsic moments of $^{220}$Rn and $^{224}$Ra provides additional support to the theoretical predictions of large enhancement factors.~\cite{Gaffney:2013}

The RadonEDM collaboration are focusing on potential EDM measurements with radon isotopes for several reasons.  Most importantly, precision measurements with  polarized noble gases in cells  have demonstrated the feasibility of an EDM experiment. For $^{129}$Xe, it was measured that $d=0.7\pm 3.4\times 10^{-27}$ \ecm~\cite{rf:Rosenberry2001}. A number of techniques have been developed including spin-exchange-optical-pumping (SEOP) using rubidium, construction of EDM cells and wall coatings that reduce wall interactions, in particular for spin greater than 1/2. The half-lives of $^{221/223}$Rn are of order 20-30 minutes, so an on-line experiment at an isotope production facility is essential. The proposed experiment (S-929) at TRIUMF's ISAC, an on-line isotope separator-facility, has been approved with high priority. The experimental program includes development of on-line techniques including collection of rare-gas isotopes and transfer to a cell, optical pumping and  techniques for detection of spin precession based on gamma-ray anisotropy, beta asymmetry and laser techniques.

For polarized rare-isotope nuclei, the excited states of the daughter nucleus populated by beta decay are generally aligned, leading to a $P_2(\cos\theta)$ distribution of gamma-ray emission. The gamma anisotropy effect has been used to detect nuclear polarization in $^{209}$Rn and $^{223}$Rn \cite{rf:Tardiff2008, rf:Kitano1988}. At TRIUMF, the large-coverage HPGe gamma-detector array TIGRESS or the new GRIFFIN array may be used. Alternatively, beta asymmetry can be used to detect nuclear polarization with a higher efficiency. Both the gamma-anisotropy and beta-asymmetry detection techniques have analyzing power expected to be limited to 0.1-0.2. The sensitivity of the EDM measurement is proportional to analyzing power, thus laser-based techniques are also under investigation. The collaboration is currently developing two-photon magnetometry for $^{129}$Xe that may also be useful as a co-magnetometer in neutron-EDM measurements. The analyzing power for two-photon transitions can be close to unity as long as the density is sufficient.

EDM measurements in radon isotopes will ultimately be limited by production rates. Fragmentation can produce useful quantities of these isotopes for development, and the beam-dump at FRIB may be a source for harvesting large quantities for an EDM measurement. Isotope-separator techniques, such as those used at TRIUMF and ISOLDE, have direct yields that are much higher, and would be a great advantage for the future of the RadonEDM program.

\subsection{Radium-225 Atomic EDM}
\label{sec:RaEDM}

The primary advantage of $^{225}$Ra is the large enhancement \cite{rf:Spevak1997, Ban:2010, Doba:2005, Gaffney:2013}, approximately a factor of 1000, of the atomic EDM  over $^{199}$Hg that arises from both the octupole deformation of the nucleus and the highly relativistic atomic electrons. This favorable case is being studied at both Argonne National Laboratory \cite{Guest:2007} and Kernfysisch Versneller Instituut (KVI) \cite{De:2009}. The scheme at Argonne is to measure the EDM of $^{225}$Ra atoms in an optical dipole trap (ODT) as first suggested in Ref. \cite{Romalis:2001}. The ODT offers the following advantages:  $\vec{v}\times\vec{E}$ and geometric phase effects are suppressed, collisions are suppressed between cold fermionic atoms, vector light shifts and parity mixing induced shifts are small.  The systematic limit from an EDM measurement in an ODT can be controlled at the level of 10$^{-30}$\ecm~\cite{Romalis:2001}.

The Argonne collaboration demonstrated the first magneto-optical trap (MOT) of Ra atoms~\cite{Guest:2007}, the transfer of atoms from the MOT to the ODT with an efficiency exceeding 80\%, the transport of atoms to an ODT in a measurement chamber 1 m from the MOT ~\cite{Parker:2012}, and spin precession of $^{225}$Ra atoms in the measurement chamber.  In the near future, they aim to begin the first phase of the EDM measurement at the sensitivity level of 10$^{-26}\ecm$, which should be competitive with 10$^{-29}\ecm$ for $^{199}$Hg in terms of sensitivity to T-violating physics. For phase 2 of this experiment, the collaboration plans to upgrade the optical trap.  In the present MOT, the slower and trap laser operate at 714 nm where there is a relatively weak atomic transition rate.  In phase 2, they would upgrade the trap to operate at 483 nm where a strong transition can be exploited for slowing and trapping.  

In Phase 1\&2, a typical experimental run will use 1-10 mCi of $^{225}$Ra presently available.  The next-generation isotope facility, such as FRIB after upgrade or Project X, is expected to produce more than 10$^{13}$ $^{225}$Ra atoms/s \cite{px2009}.  In this case it should be possible to extract more than 1 Ci of $^{225}$Ra for use in the EDM apparatus.  This would lead to a projected sensitivity of $10^{-28}-10^{-29}\ecm$ for $^{225}$Ra, competitive with $10^{-31}-10^{-32}\ecm$ for $^{199}$Hg.

\subsection{Electron EDM with polar molecules}
\label{sec:moleculeEDM}
\paragraph{\bf YbF:} Although the Standard Model predicts that the EDM of the electron is far too small to detect, being some eleven orders of magnitude smaller than the current experimental sensitivity, many extensions of the Standard Model naturally predict much larger values of eEDM that should be detectable. This makes the search for eEDM a powerful way to search for new physics and constrain the possible extensions. Cold polar molecules YbF have been used to carry out this search, setting the upper limit at $d_{\sub{e}}<1.05\times10^{-27}\,\ecm$~(90\,\% C.L.) \cite{Hudson:2011}. One should emphasize, however that the paramagnetic systems such as YbF depend strongly on two time-reversal violating (and parity-violating) interactions: the electron EDM interaction with the internal molecular electric field and a semi-leptonic scalar (quark) $\times$ pseudo scalar (electron) interaction. Since the eEDM is a dipole operator, it generically carries a power of the electron Yukawa coupling, while the scalar-pseudoscalar semileptonic interaction does not. Thus, all other inputs being equal, paramagnetic systems are more powerful probes of any physics responsible for a semileptonic system than of the eEDM. In a model independent analysis, these EDM searches constrain a linear combination of the two interactions (for a recent discussion, see {\em e.g.} Ref.~\cite{Engel:2013lsa} and references therein).

Previous paramagnetic system EDM measurements were performed on neutral heavy atoms such as Tl~\cite{Regan:2002}. Dipolar molecules have two great advantages over atoms. First, at a modest operating electric field the interaction energy of YbF due to eEDM is 220 times larger than that obtained using Tl in a much larger electric field. Second, the motional magnetic field, a source of systematic error that plagued the Tl experiment, has a negligible effect on YbF. Because of these advantages, it is possible to improve on the Tl experiment by using YbF molecules, even though the molecules are produced in much smaller numbers. The collaboration is developing a cryogenic source of YbF that yields a higher flux of molecules at three times slower velocity. With this new source, the eEDM sensitivty is likely to be pushed down to $10^{-28}\,\ecm$. Long-term plan aims to reach $10^{-30}\,\ecm$ with the development of a molecular fountain based on laser cooling of YbF.

\paragraph{\bf ThO:} The Advanced Cold Molecule EDM (ACME) collaboration uses a newly-developed cryogenic technique for creating molecular beams of unprecedented brightness~\cite{Hutzler:2011}, hence allowing large improvements in statistical sensitivity to an eEDM. ACME studies thorium monoxide (ThO), which combines the most favorable features of species used in other experiments~\cite{Vutha:2010}. In particular, the measurement takes place in the metastable H $^{3}\Delta_{1}$ state of ThO; here the effective electric field acting on the eEDM is the largest known (104 GV/cm). This state has $\Omega$-doublet substructure, which makes it possible to spectroscopically reverse the internal E-field within the molecule; this in turn enables powerful methods for rejecting most anticipated systematic errors. Finally, in the H $^{3}\Delta_{1}$ state there is a near-perfect cancellation of magnetic moments due to spin and orbital angular momenta; the resulting small magnetic moment ($< 0.01$ Bohr magnetons) makes the experiment insensitive to systematic errors and noise due to uncontrolled magnetic fields.

Based on the recent data, the collaboration has set the most stingent upper limit on eEDM: $d_{\sub{e}}<8.7\times10^{-29}\,\ecm$~(90\,\% C.L.) \cite{ACME:2013}, again assuming to scalar-pseudoscalar interaction. In the longer term, the collaboration has identified a host of methods to improve the molecular beam flux and the efficiency of state preparation and detection.  Overall, the collaboration projects a sensitivity that could ultimately reach $3\times10^{-31}$ \ecm for the eEDM.

\subsection{The Role of Theory}
\label{sec:theory}

The observation of a non-vanishing EDM in any of the aforementioned searches would constitute a major discovery in fundamental physics. Elucidating the origin of a non-vanishing EDM would require significant additional input from theoretical investigations as well as additional EDM results from searches using complementary systems. In the language of effective theory, a non-vanishing EDM of a nucleon, nucleus, paramagnetic, or diamagnetic system could arise from either the CPV QCD \lq\lq $\theta$" term or any one of twelve \lq\lq dimension six" CPV interactions involving the first generation quarks, leptons, gluons and photons. 

Unpacking the EDM search results to draw implications about these possible \lq\lq CPV sources" and their consequences for the cosmic baryon asymmetry is one of the primary challenges for EDM theory (for recent reviews, see Refs.~\cite{Engel:2013lsa,Pospelov:2005pr}). An equally important task is refining the computations of matrix elements of these sources in strongly interacting systems such as the nucleon as well as the many-body systems of atomic nuclei. At present, the theoretical uncertainties associated with these computations is significant, leading to as much as an order of magnitude or more variation in the possible magnitude of an EDM induced by any of the underlying sources\cite{Engel:2013lsa}. Lattice QCD efforts have largely focused on computing the $\theta$-term matrix elements, and even for this single source of CPV considerable room for improvement remains. The situation with diamagnetic EDMs generated from the nuclear Schiff moment is equally ripe for progress. Thus, in order to arrive at a robust conclusion regarding the possible existence of strong CPV or beyond the Standard Model \lq\lq weak scale" CPV that could generate the baryon asymmetry, one cannot rely on experiment alone. Investments in the related theoretical work are an essential part of the overall scientific effort.

\section{Weak Decays}
\label{sec:Weak}

The weak decays of light quark systems have an illustrious history in nuclear physics and have yielded fundamental insights into the nature of the electroweak interaction. Here, were review opportunities for on-going studies of pion leptonic decays, nuclear $\beta$-decay, and neutron decay. The emphasis is largely on experimental opportunities, though there have been significant theoretical developments as well. Recent reviews of the latter are available in Refs.~\cite{ Cirigliano:2013xha,RamseyMusolf:2006vr,Erler:2004cx}.

\subsection{Pion Leptonic Decay \& Electron-Muon Universality}
\label{sec:Pion}
 Electron-muon universality, within the context of the SM, is the hypothesis
that charged leptons have identical electroweak gauge interactions and differ only in
their masses and coupling to the Higgs. However, there could be additional BSM
effects, such as non-universal gauge interactions, or scalar or pseudoscalar bosons with
couplings not simply proportional to the lepton masses, that would
violate universality.

One of the most sensitive approaches to seeking such new interactions is the study of the
ratio of decay rates of pions \cite{Bryman2011}
\begin{equation}
R^\pi_{e/\mu} \equiv {\Gamma(\pi \rightarrow e \nu (\gamma)) \over \Gamma(\pi \rightarrow \mu \nu (\gamma))}
\end{equation}
which in the SM is predicted to be 1.2351(2)with an uncertainty of  $\pm$ 0.02\% \cite{Kinoshita1959,
Marciano1976,Cirigliano2007}.  BSM physics at scales as high as 1000 TeV can be
constrained or conceivably unveiled by improved measurements of this ratio. One example \cite{Bryman2011}
is a charged physical Higgs boson with couplings $g \lambda_{ud}/2 \sqrt{2}$ to the 
pseudosalar current $\bar{u} \gamma_5 d$ and $g \lambda_{l \nu}/2 \sqrt{2}$ to the current
$\bar{l}(1-\gamma_5)\nu_l$, $l=e,\mu$, where $g$ is the SU(2)$_L$ gauge
coupling and $\lambda$ represents chiral breaking suppression factors. One finds
\begin{equation}
m_{H^\pm} \sim 200 \mathrm{~TeV}~\sqrt{\lambda_{ud} (\lambda_{e \nu}-{m_e \over m_\mu} \lambda_{\mu \nu})}
\end{equation}
If $\lambda_{e \nu}/\lambda_{\mu \nu} = m_e/m_\mu$, as in the minimal 2-Higgs doublet model, 
there is no sensitivity to new physics. However, in more general multi-Higgs models
such a chiral relationship is not necessary and $\lambda$ may not be too suppressed. For example,
in the case of loop-induced charged Higgs couplings $\lambda_{e \nu} \sim \lambda_{\mu \nu} \sim 
\lambda_{ud} \sim \alpha/\pi$, one finds a $R^\pi_{e/\nu}$ determination to $\pm$ 0.1\% is sensitive to
$m_{H^\pm} \sim 400$ GeV. If a discrepancy between theory
and experiment is found in $R^\pi_{e/\mu}$, some type of charged Higgs explanation would be quite
natural.  However, such a result could also point to additional charged axial-vector interactions,  loops involving supersymmetry particles, or leptoquark-type interactions as well as R-parity violating sparticle exchange\cite{RamseyMusolf:2006vr}.
It could also be interpreted as the damping one of the $\pi_{l2}$ modes by heavy neutrino
mixing.  Analogous K  decays,
\begin{equation}
R^K_{e/\mu} \equiv {\Gamma(K \rightarrow e \nu (\gamma)) \over \Gamma(K \rightarrow \mu \nu (\gamma))},
\end{equation}
can also probe high scales and have the
added appeal of being particularly sensitive to the lepton-flavor-violating decay  $K^+ \rightarrow e^+ + \nu_\tau$,
a process that might be induced through loop effects \cite{Masiero2006}.\\

\subsubsection{Experimental Studies of $\pi \rightarrow e \nu$ and $K \rightarrow e \nu$ Decays}
The most recent $\pi^+ \rightarrow e^+ \nu$ ($\pi_{e2}$) branching ratio measurements and subsequent
determination of the ratio $R^\pi_{e/\mu}$ were done at TRIUMF \cite{Britton1992} and PSI \cite{Czapek1993}
in the 1990s.  The
results from the two experiments are consistent and in  good agreement with the SM
expectation previously discussed,
\begin{equation}
R^{\pi-\mathrm{TRIUMF}}_{e/\mu} = 1.2265(34)(44) \times 10^{-4} ~~~~~R^{\pi-\mathrm{PSI}}_{e/\mu} = 1.2346(35)(36) \times 10^{-4},
\end{equation}
where the first and second uncertainties are due to
statistical and systematic effects. The PDG average value is
$R^{\pi}_{e/\mu} = 1.230(4) \times 10^{-4}$ \cite{Nakamura2010},
which includes results from \cite{Bryman1986}. Two new experiments are underway 
at TRIUMF \cite{Aguilar2009} and PSI \cite{Pocanic2004} which promise to
improve the precision of $R^{\pi}_{e/\mu}$ by a factor of 5 or more, thereby
testing the SM prediction to better than 0.1\%. At that level of precision, new physics effects could
appear as a deviation, or in the absence of a deviation,
strong new
constraints would be imposed on such physics.\\

The PIENU experiment \cite{Aguilar2009} is based on a refinement of the technique used in the previous TRIUMF
experiment \cite{Britton1992}. The branching ratio will be obtained from the ratio of positron yields from
the $\pi \rightarrow e \nu$ and from the $\pi \rightarrow \mu \rightarrow e$ decay chain. By measuring 
the positrons from these decays
in a non-magnetic spectrometer, many normalization
factors, such as the solid angle of positron detection, cancel to first order, so that only small corrections for 
energy-dependent effects, such as those for multiple Coulomb scattering (MCS) and
positron annihilation, remain. Major improvements in precision stem from
the use of an improved geometry, a superior calorimeter, high speed digitizing of all pulses,
Si strip tracking, and higher statistics. The improvements lead to an expected precision on
the $R^{\pi}_{e/\mu}$  branching ratio of $<0.06$ \%, which corresponds to a 0.03 \% uncertainty in the ratio
of the gauge boson-lepton coupling constants $g_e/g_\mu$ testing electron-muon universality.\\

At PSI, the PIBETA CsI spectrometer \cite{Pocanic2004}, built for a determination of the $\pi^+ \rightarrow \pi^0e\nu$
branching ratio and other measurements \cite{Frlez2004}, has been upgraded and enhanced for the
PEN \cite{Pocanic2009} measurement of the $\pi \rightarrow e \nu$ branching ratio. The PEN 
technique is similar to that
employed in the previous PSI experiment \cite{Czapek1993}, which used a nearly 4$\pi$-sr BGO spectrometer.
PEN began operations in 2007 and has observed $>10^7$ $\pi \rightarrow e \nu$ decays. PEN completed
data acquisition 2010 and expects to reach a precision of $<$ 0.05\% in $R^\pi_{e/\mu}$. \\

Recent progress on $R^K_{e/\mu}$ 
has been made by KLOE \cite{Ambrosino2009} and NA62 \cite{Goudzovski2009},  with current NA62 efforts
aimed at reaching a precision of  0.4\%. KLOE collected 3.3 billion $K^+K^-$
pairs, observing decay products in a drift chamber in a 0.52T axial magnetic field
surrounded by an electromagnetic calorimeter. The measurement of $R^K_{e/\mu}$ consisted of comparing the
corrected numbers of decays observed from the $K \rightarrow e \nu(\gamma)$ and $K \rightarrow \mu \nu(\gamma)$ channels. The result,  $R^{K~KLOE}_{e/\mu} = (2.493 \pm 0.025(\mathrm{stat}) \pm 0.019(\mathrm{syst})) \times 10^{-5}$ \cite{Ambrosino2009},
agrees with the SM
prediction at
the 1\% level.\\

NA62 at CERN using the setup from NA48/2 has embarked on a series of $K_{e2}/K_{\mu 2}$
measurements \cite{Goudzovski2009}. The $K^+$ beam is produced by the 400 GeV/c SPS. Positively charged
particles within a narrow momentum band of (74.0 $\pm$ 1.6) GeV/c are selected by an
achromatic system of four dipole magnets and a muon sweeping system, enter a fiducial
decay volume contained in a 114 m long cylindrical vacuum tank producing a secondary
beam. The $K \rightarrow e \nu(\gamma)$ and $K \rightarrow \mu \nu (\gamma)$  detection system includes a magnetic
spectrometer, a plastic scintillator hodoscope, and a liquid krypton
electromagnetic calorimeter. As in KLOE, the experimental strategy is based on
counting the numbers of reconstructed $K \rightarrow e \nu(\gamma)$ and $K \rightarrow \mu \nu (\gamma)$ events concurrently,
eliminating dependence on the absolute beam flux and other potential systematic
uncertainties. The result,  $R^{K~NA62}_{e/\mu} = (2.487 \pm 0.011(\mathrm{stat}) \pm 0.007(\mathrm{stat})) \times 10^{-5}$ \cite{Goudzovski2009},
is based on 40\% of the data acquired in 2007 and agrees with the SM prediction.
The full data sample may allow a statistical uncertainty of 0.3\% and a total
uncertainty of 0.4-0.5\% .\\

\subsubsection{Future Prospects}
If PIENU and PEN achieve their sensitivity goals there will still be a considerable window
in which new physics could appear without complications from SM prediction uncertainties. To reach
the uncertainty ~0.02\%, the level  of current SM calculations, would require
a new generation of experiments capable of controlling
systematic uncertainties at or below 0.01\%. High-precision measurements of
$R^{\pi/K}_{e/\mu}$ will be an important complement to LHC high-energy studies. 
High intensity beams with 100\% duty factors and ultra-high intensities
and purities will be important to ultra-high precision experiments on $\pi/K \rightarrow e/\mu$.
Such experiments potentially could lead to breakthroughs in our understanding of $e-\mu$ universality and
are sensitive to a variety of subtle non-SM effects.
Project X would provide such beams for pions and kaons, while
pion studies could continue with the beams available at PSI or TRIUMF.

\subsection{ Nuclear and Neutron Decay: CKM Unitarity and Non-Standard Interactions}
\label{sec:Super}

Studies of weak interactions based on nuclear $\beta$ decay are currently focused on
probing the limits of the Standard Model: Both the conserved vector current (CVC)
hypothesis and the unitarity of the Cabibbo-Kobayashi-Maskawa (CKM) matrix are being
tested with ever increasing precision. To do so, it is necessary first to isolate the
vector part of the combined vector and axial-vector ($V$$-$$A$) structure of the weak
interaction, a requirement that is satisfied by superallowed $0^+$$\rightarrow 0^+, \Delta T = 0$
nuclear decays, which are pure vector transitions.
These transitions occur in a wide range of nuclei (10$\le A \le$74) and yield
the vector interaction strength in each case.  If CVC is valid, then the
strength of the vector interaction is not renormalized in the nuclear medium but is a
`true' constant; and, so far, the constancy of the strength is confirmed by these
superallowed decays at the level of $10^{-4}$ \cite{HT09}.  With CVC satisfied, these
results also lead to today's most precisely determined value of the CKM matrix element,
$V_{ud}$: $0.97425 \pm 0.00022$ \cite{HT09}.

When $V_{ud}$ is combined with the $V_{us}$ and $V_{ub}$ values obtained from kaon and B-meson decays, 
respectively, the sum of squares of the top-row elements of the CKM matrix provides the
most demanding test available of the unitarity of the matrix.  The result, currently standing
at 0.99990(60) \cite{TH10a}, agrees with unitarity. Its precision,
which can still be improved, yields stringent constraints on (V,A) current interactions,
with generic limits on the effective scale for new physics at roughly the 10 TeV level.\cite{Ciri10}
A large assortment of extensions to the standard model, including new Z' gauge bosons,
generic Kaluza-Klein W* excitations, and charged Higgs bosons, are tightly
constrained by the unitarity sum.\cite{Anto10}.  Flavor universality in supersymmetric
extensions of the standard model are also constrained by the unitarity sum.\cite{Kury02}
The robustness of these limits and enormous progress made in the Kaon-sector
in the precision and reliability with which $V_{us}$ can be determined motivate
continued effort on the experimental extraction of $V_{ud}$.\cite{Anto10}

Other tests of the weak interaction involve asking whether it is pure $V$$-$$A$ as assumed in the Standard Model or
whether there are small components of $S$ (scalar) or $T$ (tensor) interactions.  The
$0^+$$\rightarrow 0^+$ superallowed decays also set a limit on $S$ but not on $T$. The study of $\beta$ decay between ($J^\pi, T$) = ($0^+, 1$) nuclear analog states has
been a fertile means of testing the Standard Model.  Because the axial current cannot
contribute to transitions between spin-0 states, only the vector current is involved in
these transitions.  Thus, according to CVC, the experimental $ft$-value for each of these
superallowed transitions should be simply related to the fundamental weak-interaction
coupling constant, $G_V$.  The $ft$-value itself depends on three measured quantities: the
total transition energy, $Q_{EC}$, the half-life, $t_{1/2}$, of the parent state, and the
branching ratio, $R$, for the particular transition of interest.  The first of these can
be measured in a Penning trap, where very few ions are sufficient for high precision;
however, both $t_{1/2}$ and $R$ require large numbers of observed decays to achieve the
required statistics.
Currently, the best-known superallowed decays are from nuclei rather close to stability,
which are easily produced. Future improvements in precision will need
comparable measurements on transitions from nuclei much farther from stability.  Higher
intensity beams will be required to produce them.

Neutron $\beta$ decay is conceptually the simplest example of a mixed vector and axial-vector
$T = 1/2$ mirror decay. The neutron lifetime depends on the combination of vector and axial vector couplings
$g_V^2+ 3g_A^2$. In order to obtain a precise value of $V_{ud}$ one must separately determine the ratio 
 $\lambda = g_A/g_V$ through the measurement of a neutron decay correlation. Doing so then allows one to 
 extract $V_{ud}$ from the value of $g_V$ without encountering the QCD uncertainties associated with $g_A$. While this strategy is
 conceptually simple the experimental challenges encountered in confining the neutron
long enough to measure its properties are non-trivial.  At present there are conflicting results -- well outside of statistics --
for the neutron lifetime.  The internal agreement among the correlation measurements is better, but
is far from satisfactory since there has been a systematic drift in the measured central value of $\lambda = g_A/g_V$
over the past two decades.  More measurements are clearly required. In the following sections we survey the prospects in more detail.

There are two theoretical roadblocks to  quest for a more accurate
$V_{ud}$ value.  First, there is a radiative correction to be included
in the analysis, and a part of this, known as the `inner' radiative 
correction, is not well determined.  A recent evaluation by Marciano
and Sirlin \cite{MS06} reduced its uncertainty by a factor of two;
but, even so, the new uncertainty still remains the largest contributor
to the error budget for $V_{ud}$.  Second, the use of the CVC hypothesis
is valid only in the isospin-symmetry limit.  In nuclei, the presence
of the Coulomb force acting between protons breaks isospin symmetry as
does charge dependence in the nuclear force to a lesser extent.  So
an isospin-symmetry breaking correction, denoted $\delta_C$, needs
to be evaluated and this involves a nuclear-structure-dependent
calculation.  The estimated uncertainty in the model dependence of
these calculations is the second largest contributor to the error
budget for $V_{ud}$.  There has been a lot of recent activity
in $\delta_C$ calculations for the $0^+$$\rightarrow 0^+$ superallowed
decays \cite{Sa11,Gr10,MS08,Au09,LGM09,TH08}, with considerable
disparity being evident among the results.

Fortunately, there is an opportunity here.  Towner and Hardy \cite{TH10b}
have proposed an experimental test rooted in the requirement that the
$\delta_C$ calculations should yield results consistent with a 
conserved vector current.  To date, shell-model-based calculations \cite{TH08}
satisfy this test the best and it is these calculations that have been
used in the extraction of $V_{ud}$.  Further precise experiments could
improve the test and reduce the uncertainty on $V_{ud}$.  Calculations
anticipate somewhat larger $\delta_C$ values for the decays of $T_z$\,=\,-1
superallowed emitters with $A \leq 42$, and much larger values for $T_z$\,=\,0
emitters with $A \geq 62$.  If experiment confirms these large calculated
$\delta_C$ values, then it validates the much smaller values for the transitions
now used to obtain $V_{ud}$. In particular, the different calculations give a
range of predictions for mirror pairs of transition such as: $^{26}$Si ($T_z$\,=\,-1)
$\rightarrow$ $^{26}$Al ($T_z$\,=\,0) versus $^{26}$Al ($T_z$\,=\,0) $\rightarrow$
$^{26}$Mg ($T_z$\,=\,1).  The decay of the $T_z = -1$ member of each pair requires a very difficult
branching-ratio experiment to be performed but, if successful, the result would be
very revealing for the $\delta_C$ calculations.  For the other group of emitters with $A \geq 62$,
only $^{62}$Ga has so far been measured with the requisite precision.  Experiments on
lifetimes, branching ratios, and $Q$-values for $^{66}$As, $^{70}$Br and $^{74}$Rb
would be very welcome.  

A recent bellwether experiment \cite{Me11} focused on the beta decay of $^{32}$Cl.
One of its branches is a $J$$\rightarrow$$J$, $\Delta T$\,=\,0 transition for which theory
fortuitously predicts a negligibly small axial-vector component.  Thus it could be
analysed as if it were a pure Fermi transition.  The result it yielded corresponded
to a very large isospin-symmetry-breaking correction of order $5 \%$.  Being a nucleus with
$A = 4n$, large isospin-symmetry breaking could be anticipated since the daughter state
lies very close in energy to a state of the same spin and different isospin.  As a consequence,
the case provides a critical challenge to the $\delta_C$ calculations, a challenge which was successfully
met by the same type of shell-model-based calculations used to analyze the $0^+$$\rightarrow 0^+$
transitions.  In future an examination of other $J \rightarrow J$
transitions from $A = 4n$ nuclei should be undertaken to seek other examples with small
axial-vector contributions accompanied by large isospin-symmetry breaking.

%%%%%%%%%%%% Neutron Decay %%%%%%%%%%%%

\subsection{Neutron Decay}
\label{sec:nDecay}

Measurements of neutron beta-decay  provide basic
parameters for the charged weak current of the
nucleon.\cite{Abel08}  In particular, neutron beta-decay measurements are the definitive
source for $g_A$, the axial form factor, and provide a nuclear-structure-independent value
for $V_{ud}$, the CKM matrix element associated with ud quark currents.  Although at
present the $0^+ \rightarrow 0^+$ superallowed decays provide the most precise value
for $V_{ud}$, the experimental data for the neutron continues to improve, and should
become directly competitive with $0^+ \rightarrow 0^+$ during the next ten years.

We note that $g_A$ is important in our understanding of
the spin and flavor structure of the nucleon\cite{Bass05,Clos88}, a central target
for high precision lattice QCD studies\cite{Yama08,Choi10}, an essential parameter
for effective field theories\cite{Gock05}, and one of a small set of parameters
necessary in establishing high precision predictions of solar fusion\cite{Adel10}.
The neutron lifetime figures prominantly in high precision
predictions for big bang nucleo-synthesis as well\cite{Burl99}.  High precision values
for $g_A$ are also important for the reactor neutrino-anomaly question, one
of the results driving current interest in short baseline oscillations
studies\cite{Ment11}.  The current value of $g_A$ is $g_A$ = 1.2701(25)\cite{Naka10}.

High precision neutron beta-decay studies also provide constraints on a large 
variety of extensions to the standard model.  The most stringent
constraints come through tests of the quark-lepton universality
of the weak interaction, or tests of CKM unitarity as discussed above.  As also mentioned,
the precision of $V_{ud}$ from $0^+ \rightarrow 0^+$ decays is nominally limited by loop-level
electroweak radiative corrections\cite{Hard09,Marc06}, however the nuclear structure
dependent corrections for the $0^+ \rightarrow 0^+$ systems remain an area of active
concern. Neutron beta-decay can provide a structure-independent value for $V_{ud}$, a
significant contribution to the status of the current unitarity test.

The observables in neutron decay include a number of correlations (and
the Fierz term, which influences the energy dependence of the total beta-decay
rate) that provide multiple probes of non V-A interactions generated by
standard model extensions.\cite{Abel08,Gudk06}  For example, constraints on (S,T) interactions
arise from angular correlations measurements such as the neutrino-asymmetry
and the Fierz term.  Because two observables with similar sensitivities to
these terms are available, there is a consistency test within the neutron decay
system itself for these effects. In particular, it is the aim of some beta-decay
experiments (in the planning or construction phase at present) to reach sensitivities
of a few parts in $10^4$.  In this case, the model-independent constraints for
interactions which only couple to electrons and induce scalar and tensor terms can be made
quite stringent with next generation beta-decay experiments. For example,
limits in the 5-10 TeV range which are significantly stronger than expected LHC limits
are expected to be feasible.  In addition, if a new particle resonance is discovered
at the LHC, beta-decay experiments at this level of precision may provide complementary
information on the quantum numbers and weak couplings of such a resonance, as was
recently demonstrated for the case of a scalar resonance\cite{Batt11}. Relevant limits
(complementary to those placed by LHC) can also be placed
on supersymmetric couplings\cite{Prof06} and couplings to leptoquarks\cite{Seve06}. 

T-noninvariant angular correlations can also be probed in beta-decay.
These experiments can provide constraints on CP violating phases beyond the standard model
that are complementary to the ones derived from  EDMs.\cite{Herc01,Tuli11}
In particular, a number of measurements have been performed of angular correlations
proportional to complex (V,A)couplings (parameterized by the "D" coefficent). The emiT\cite{Mumm11} and TRINE\cite{Sold04} collaborations have
established the basis for pushing sensitivies for T-violating phases to the final state effect level ($10^{-5}$ level).

The past ten years have seen significant growth in the number of physicists
involved in neutron beta-decay measurements. Although it is beyond the scope
of this brief summary to catalog all of the experimental activity in this subfield,
it is characterized by often complementary experiments with cold and ultracold neutrons
and has seen the emergence of precision measurements of radiative decay of the neutron
\cite{Nico05} for the first time. A number of experiments are underway which target
precisions near or at the 0.1\% level in the next few
years for the lifetime\cite{Dewe09}, the electron-neutrino-asymmetry\cite{Sims09,Wiet09} and the
beta-asymmetry\cite{Maer09,Liu10}.  Taken as a group, they provide a powerful
consistency test for the form factors and standard model constraints which can be
extracted at this level of precision\cite{Gard01}.

Ongoing measurements have also set the stage for a number of ambitious experiments under
development or construction which target precisions in the $10^{-4}$ range.  For ultracold
neutrons, for example, there are lifetime experiments based on material and magnetic
trapping geometries\cite{Fomi11,Wals09} and angular correlation experiments under
development particularly sensitive to (S,T) interactions\cite{Wilb09}.  For angular
correlations measurements with cold neutron beams, the PERC\cite{Dubb08} collaboration
based in Munich have as their goal polarimetry and other systematic
errors ultimately in the low $10^{-4}$ range, and the Nab/ABba\cite{Poca09,Wilb05}
collaboration will be targeting systematic uncertainties below the
$10^{-3}$ level for their measurements as well.

Intensity frontier development should provide the opportunity to optimize existing
cold neutron beams delivery for fundamental neutron physics research, positively
impacting beta-decay experiments as well as a variety of other fundamental neutron
studies.  For ultracold neutron-based experiments, the intensity frontier
initiative could provide the opportunity to construct a next-generation source of
extracted ultracold neutrons.  Although work over the past ten years has established
viable strategies to significantly increase ultracold neutron densities, experiments
remain strongly constrained by the ultracold neutron densities at existing sources.
A next-generation source could permit the community to capitalize on the ongoing
refinement of systematic errors in existing beta-decay, EDM and short-range interaction
searches.  In particular, for beta-decay studies, it should enable the next generation
of beta-decay experiments with ultracold neutrons to reach sensitivities limited by
systematic errors, and probe energy scales comparable to and in some cases above that
planned for the LHC. 

%%%%%%%%%%%%%%%%%% Neutron Lifetime %%%%%%%%%%%%%%%%

\subsection{Neutron Lifetime}
\label{sec:nlife}

Measurements of the neutron lifetime have been approaching the 0.1\% level of precision ($\sim$ 1~s uncertainty). However, several recent neutron lifetime results~\cite{Serebrov05, Serebrov2008, Ezhov09, Pichlmaier10} are up to 7~s lower than the PDG value before 2010 ($885.70\pm 0.85$~s)~\cite{PDG06}.
This $>\mbox{6 }\sigma$ deviation has not yet been resolved.
Even though the new 2011 update of the PDG value ($881.5 \pm 1.5$~s)~\cite{PDG2011} includes all these measurements, with the uncertainty scaled up by a factor of 2.7, the PDG questions this new world average and calls upon the experimenters to clear up the current state of confusion.

The precise determination of the neutron lifetime could also have profound impact on astrophysics and cosmology. 
In the standard big bang nucleosynthesis (SBBN) model, the abundance of the light elements can be determined with a single cosmological parameter -- the baryon-to-photon ratio $\eta_{10}$, together with the nuclear physics input of the neutron lifetime and 11 key nuclear reaction cross-sections~\cite{Burles99}.
The influence of the neutron lifetime on the abundance of the light species of primordial nuclei, in particular $^4$He, is based on two effects. First, the neutron lifetime indirectly affects the neutrino-nucleon reaction rate and the neutron-to-proton ratio when the primordial neutrons decouple from the radiation field. Second, the lifetime directly affects the number of these neutrons left to participate in nucleosynthesis, which is delayed by photo-dissociation of deuterons in the hot radiation field. 
As the primordial neutrons are protected against $\beta$-decay by fusing with protons into deuterons and then into $^4$He, a shorter neutron lifetime would result in a smaller $^4$He abundance (Y$_p$).
As a consequence, a 1\% change in the neutron lifetime leads to a 0.75\% change of Y$_p$~\cite{Mathews05}.
With a precise determination of $\eta_{10}$ from WMAP, the SBBN predicts Y$_p$ with a 0.2 -- 0.3\% precision~\cite{Steigman10}.
Of the primordial elements, Y$_p$ is particularly sensitive to the expansion rate of the universe and to a possible lepton asymmetry in the early universe~\cite{Steigman07}.
Information on Y$_p$ is attained from either direct observations of the H and He emission lines from low-metallicity extragalatic regions or from the indirect measurements using the power spectrum of the cosmic microwave background (through its effect on the electron density at recombination). The current precision of these measurements is about 1 -- 2\%~\cite{Steigman10, Izotov2010}. 
With the anticipated improvement from the Planck experiment in Europe and the James Webb space telescope in US, the interpretation of Y$_p$ in SBBN will hinge on the accuracy of the neutron lifetime.   
In astrophysics, the weak axial coupling constant $g_A$ of neutron $\beta$-decay is a key parameter in the standard solar model~\cite{Bahcall82}, which describes the $pp$ fusion process and the CNO cycle that produce solar energy as well as generate the solar neutrinos that can be directly measured by terrestrial experiments. 
Solar neutrino fluxes thus depend indirectly on the neutron lifetime and its uncertainties.
As more measurements of the solar neutrino flux become available~\cite{Borexino08, SNO}, the precise knowledge of the neutron lifetime will have a greater impact on the expected values. 
The same concern also applies to the estimate of the anti-neutrino flux used in many neutrino oscillation experiments based at reactors~\cite{Mention11}.

It seems that the only way to settle the controversy surrounding the value of the neutron lifetime is to perform independent measurements with 0.1~s precision and rigorous control of systematic effects.
The difficulties with measuring the absolute neutron lifetime originate from the low energy of its decay products, the essential impossibility of tracking slow neutron trajectories in matter, and the fact that the lifetime is long.
The lifetime of $\beta$-decay is comparable to the time-scale of many surface effects that contribute to the loss of neutrons~\cite{Lamoreaux02, Steyerl10}. To achieve a precision measurement of the $\beta$-decay lifetime, one has to control these additional sources of loss to levels better than the desired precision.
Recent lifetime measurements, which succeeded in reducing the uncertainty to a few seconds, have used ultracold neutrons (UCN) trapped in material bottles.
With small kinetic energies, UCNs experience total reflection from material walls (with a small absorption coefficient) at any incident angle~\cite{golub1991}.
This property allows them to be trapped in material bottles for times comparable to the $\beta$-decay lifetime.
In a typical UCN bottle experiment, the UCN are loaded into a bottle (or a trap) and stored for different periods of time, and the survivors are counted by dumping them into a UCN detector outside the trap. 
There is much controversy over how to reliably correct for the loss of UCN when they interact with material on trapping walls.

The UCN$\tau$ collaboration will measure the neutron lifetime using UCN in a novel magneto-gravitational trap~\cite{Walstrom09}.  
This trap will eliminate interactions on the confining walls and the associated uncertainties by replacing the material bottle with a trap formed by magnetic fields on the sides and bottom, and closed by gravity at the top.  
Since the interactions of neutrons with magnetic fields and gravity are well understood (and can be reliably modeled in numerical simulations), the systematic uncertainties in this approach are expected to be smaller than (and independent of) those in earlier experiments using material bottles.
The novelty of the experiment originates from the use of an asymmetric trap to facilitate (1) fast draining of the quasi-bound UCN~\cite{Bowman05}, and (2) quick sampling of the entire phase space in order to suppress spurious temporal variation of the decay signals when coupled to a non-uniform detection efficiency. 
The room temperature trap using a permanent magnet array avoids many engineering challenges of prior cryogenic experiments and allows fast turn-around time for detector prototyping. 
The design of the experiment, with an open top, provides ample room for implementing many novel detection techniques, allowing a comprehensive study of the systematic effects discussed above. 

%In the preliminary report of the 2011 review on fundamental neutron physics~\footnote{http://science.energy.gov/$\sim$/media/np/nsac/pdf/mtg-63011/Kumar\_Neutron\_Interim\_Report.pdf}, conducted by the National Science Advisory Committee (NSAC), the committee strongly recommends the pursuit of the planned NIST beam experiment at the 0.1\% level of accuracy, and encourages more R\&D effort on the UCN$\tau$ experiment. 

The UCN$\tau$ experiment uses techniques complementary to the beam experiment, and has the potential to push the precision of the neutron lifetime beyond the current state-of-the-art, towards the 0.01\% level.
The Intensity Frontier of Project X brings about the unique opportunity to install a world-class UCN source, driven by the planned proton source. 
This investment will add a new facet to strengthen the scientific program aimed to probe physics beyond the SM at the TeV scale, using complementary techniques with low energy neutrons, including the search for the neutron electric dipole moment, the $n\bar{n}$ oscillation, and a comprehensive program of precision $\beta$-decay measurements.  

%%%%%%%%%%%%%%%%%% Nuclear Correlations %%%%%%%%%%%%%%%

\subsection{Nuclear Decay Correlations}
\label{sec:nuccor}

Two tests of discrete symmetries are being considered at the National Superconducting Cyclotron Laboratory, Michigan State University, via
measurements of correlation terms in nuclear beta decay. The first
project plans a differential measurement of the so-called
polarization-asymmetry correlation in the decay of $^{21}$Na as a tool
to search for deviations from maximal parity violation. The second
is the measurement of a fifth-fold correlation in $^{36}$K decay that
is sensitive to deviations from time reversal invariance. Both
measurements require spin polarized nuclei that can be produced at
the laser spectroscopy and beam polarization facility, BECOLA. 
Low energy (maximum 60 keV/q) polarized nuclei will be produced by
first stopping the high energy fragments with suitable "stoppers" and
then by transporting the thermal beams towards the BECOLA beam line
where a collinearly overlapped laser light induces their polarization
by the optical pumping technique.

%%%%%%%%%%%%%%%%%%%%%%%%%%%%%%%%%%%%%%%%%%%%%%%%%%%%%%%%%%%%%%%%%%%%%%%%%%%%%%%
\subsubsection{The differential polarization-asymmetry correlation}

Measurements of pseudo-scalar quantities in beta decay can probe
possible deviations from maximal parity violation due for instance to
the presence of right-handed bosons with vector and axial couplings or
to the exchange of other bosons with exotic scalar and tensor couplings.

The most stringent tests of maximal parity violation in nuclear beta
decay arise from measurements of polarization-asymmetry correlations
in the decays of $^{107}$In \cite{Sev93} and $^{12}$N \cite{Tho01}.
In comparison with the beta asymmetry parameter and with the longitudinal
beta polarization, this observable offers an enhanced sensitivity to
deviations from maximal parity violation resulting from the combination
of two pseudo-scalar quantities contributing to the decay.
In previous experiments, the nuclear polarization was obtained by
either a low temperature nuclear orientation technique or by
polarization-transfer in a reaction initiated with a polarized beam.
All measurements so far have been carried out at fixed beta particle
energies by selecting a window in the beta energy spectrum with magnetic
spectrometers.

The longitudinal polarization of beta particles emitted from unpolarized
nuclei has a constant sign and also a constant sensitivity to effects
related to partial parity symmetry restoration. Beta particles emitted
from polarized nuclei can in contrast exhibit a longitudinal polarization
that changes sign as a function of the beta particle energy. In addition,
the sensitivity to effects associated with parity symmetry restoration
increases at lower beta particle energies. This offers a new window
for tests of maximal parity violation in beta decay provided the SM
values can be controlled to sufficient accuracy.

A new positron polarimeter is being designed for a differential
measurement of the longitudinal polarization of beta particles emitted
from polarized $^{21}$Na nuclei.

%%%%%%%%%%%%%%%%%%%%%%%%%%%%%%%%%%%%%%%%%%%%%%%%%%%%%%%%%%%%%%%%%%%%%%%%%%%%%%%
\subsubsection{Search for time reversal violation by a fifth-fold correlation measurement}

The five-fold correlation
$E_1\vec{J}\cdot(\vec{p}\times\vec{k})(\vec{J}\cdot\vec{k})$,
where $\vec{J}$ is the nuclear spin, $\vec{p}$ is the beta particle
momentum, and $\vec{k}$ is the photon momentum provides a mean
to the searches for time reversal violation \cite{Mor57, Cur57, Hol72}
that is complementary to measurements of triple correlations in beta decay.
The fifth-fold correlation is P-odd/T-odd and it is generally interpreted
in terms of a possible imaginary phase between the vector and axial
couplings. The most precise result so far for such a correlation was
obtained with $^{56}$Co nuclei and yielded $E_1 = -0.011\pm0.22$ \cite{Cal77}.
This provides the weakest constraint compared with the other T-violating coefficients
in beta decay. The transition of interest here is an isospin hindered
Gamow-Teller transition. The contribution from the Fermi matrix element
is finite, though small, due to a breakdown of isospin symmetry or possibly
through a contribution of second class currents (SCC). If SCC --which
are known to be zero so far \cite{Wil00}-- were the source for the underlying
mechanism for T-violation, the $^{56}$Co decay provides an ideal test. However,
no $E_1$ test has been performed so far in a mirror beta decay.

It has been pointed out \cite{You95} that the superallowed decay of
$^{36}$K is a good candidate to test T symmetry by a fifth-fold correlation
measurement.

The low energy, polarized $^{36}$K beam will be implanted in a host
crystal placed under a magnetic field surrounded by a set of high resolution
germanium detectors. The decay of interest populates a $2^+$, $T = 1$ state
at 6.61 MeV in the $^{36}$Ar daughter with a branching ratio of 42\%.
The excited state in $^{36}$Ar decays to the ground state by emission of
several gamma rays with energies larger than 2 MeV providing the conditions
needed for a $\beta\gamma$ angular correlation experiment. 

To be competitive with present limits, the new T-invariance measurement
will most likely require the higher beam rates like those expected at the
future FRIB facility at MSU.

\subsection{$\beta$-Decay with Neutral Atom Traps}
\label{sec:trapsI}

Recent advances in the techniques of atom and ion trapping have opened up a new 
vista in precision $\beta$ decay studies due to the near-textbook source they 
provide: very cold ($\lesssim1$~mK) and localized ($\lesssim1$~mm$^3$), with 
an open geometry where the daughter particles escape with negligible 
distortions to their momenta.  

Magneto-optical traps (MOTs) have demonstrated the ability to measure the 
angular distribution of short-lived radioactive neutral atoms. 
Experiments using MOTs have placed stringent direct limits on a possible 
fundamental scalar current in the charged weak interaction via a precise 
measurement of the $\beta-\nu$ correlation parameter, 
$a_{\beta\nu}$~\cite{gorelovPRL,scielzo}. In a beta decay, the parent nucleus produces three products: a $\beta^+$, a $\nu$, and a recoiling daughter nucleus, which freely escapes the trap. By detecting it in coincidence with 
the $\beta^+$, both the momentum and angular distribution of the
$\nu$ can be measured~\cite{behr}. 

Techniques are  being developed by the \trinat{} collaboration at \triumf{} to highly polarize 
laser-cooled atoms via optical pumping. 
They  have published a measurement of the $\nu$ asymmetry $B_{\nu}$ of $^{37}$K at 3\% accuracy~\cite{melconianPLB,pitcairn,melconian}.
The goals of the spin correlation program include a simultaneous 
measurement of $\beta$ and recoil asymmetries 
$A_\beta$ and $A_{\rm recoil}$ at part per thousand
sensitivity, and eventually $B_\nu$ at 0.3\%.  
E.g., a week of counting of $A_{\rm recoil}$ would have statistical sensitivity
to $C_t+C_t'$ at 0.002. 

The  collaboration would also improve their best limits on scalar interactions coupling to the first generation of particles by measuring the $\beta$-$\nu$ correlation in the pure Fermi decay of $^{38{\rm m}}$K~\cite{gorelov}. 
The complete angular acceptance for recoils will minimize key systematic errors in the upgraded experiment, with a goal of reaching 0.1\% accuracy in $a$ and possibly $b_{\rm Fierz}$.

 Concentrating on isobaric analog decays allows the determination of small
recoil-order standard model corrections from the electromagnetic moments of 
the nuclei. The Gamow-Teller/Fermi ratio in $^{37}$K can be well-determined from measurements of the lifetime and branching ratio, as the community has achieved a reliable set of techniques for these experiments.

If successful, these correlation experiments would be complementary to constraints from radiative $\pi$ decay~\cite{bhat} 
and indirect EFT-dependent constraints from $\pi$ to e $\nu$~\cite{campbell}, 
and begin to allow sensitivity to SUSY left-right sfermion mixing~\cite{profumo}.

It is natural in the trap to measure the time-reversal violating
D $ \vec{I} \cdot \vec{p_\beta} X \vec{p_\nu}$. In a conceptual design
for a dedicated geometry with 20\% $\beta$ efficiency (that the polarized light comes
at 90$^{\rm o}$ to the $\beta$ detectors helps) statistical
error/week of 2x10$^{-4}$ seems achievable.
The best measurement in the decay of the
neutron was recently published by
emiT, achieving 2x10$^{-4}$ sensitivity~\cite{emiT}.
However, a recent effective field theory calculation has identified a
mechanism for the electric dipole moment of the neutron 
to constrain leptoquark contributions to D (previously considered 
safe from EDM's~\cite{herczeg}, and if
there are no cancellations would imply D $<$ 3x10$^{-5}$~\cite{tulinD}. 
Since there are many possible contributions to CP violation from new physics,
D could still be complementary. We will study time-reversal systematics in the present
$^{37}$K geometry.

%%%%%%%%%%%%%%%% He6 Tensor Currents %%%%%%%%%

\subsection{Searching for Tensor Currents in $^6$He}
\label{sec:He6}

At the CENPA Tandem Accelerator of the University of Washington, there is a program to search for tensor currents in the decay of $^{6}{\rm He}$\cite{He6 collaboration}. The decay correlation parameter $a$ is sensitive {\em quadratically} to tensor currents of either chirality, while $b$ is sensitive {\em linearly} to tensor currents, but only to those with no left-handed anti-neutrinos (or right-handed neutrinos). The aim is to detect $a$ with a relative precision of $\sim 0.1$\% and $b$ with absolute precision of $\sim 10^{-3}$. Because of its linear dependence on the tensor couplings, $b$ is more sensitive to tensor currents. However, because it is sensitive to only one chirality, a test concentrating solely on $b$ would leave unchecked a large region of the parameter space with a particular bias. 

\begin{figure}[ht]
\includegraphics[width=0.4\textwidth]{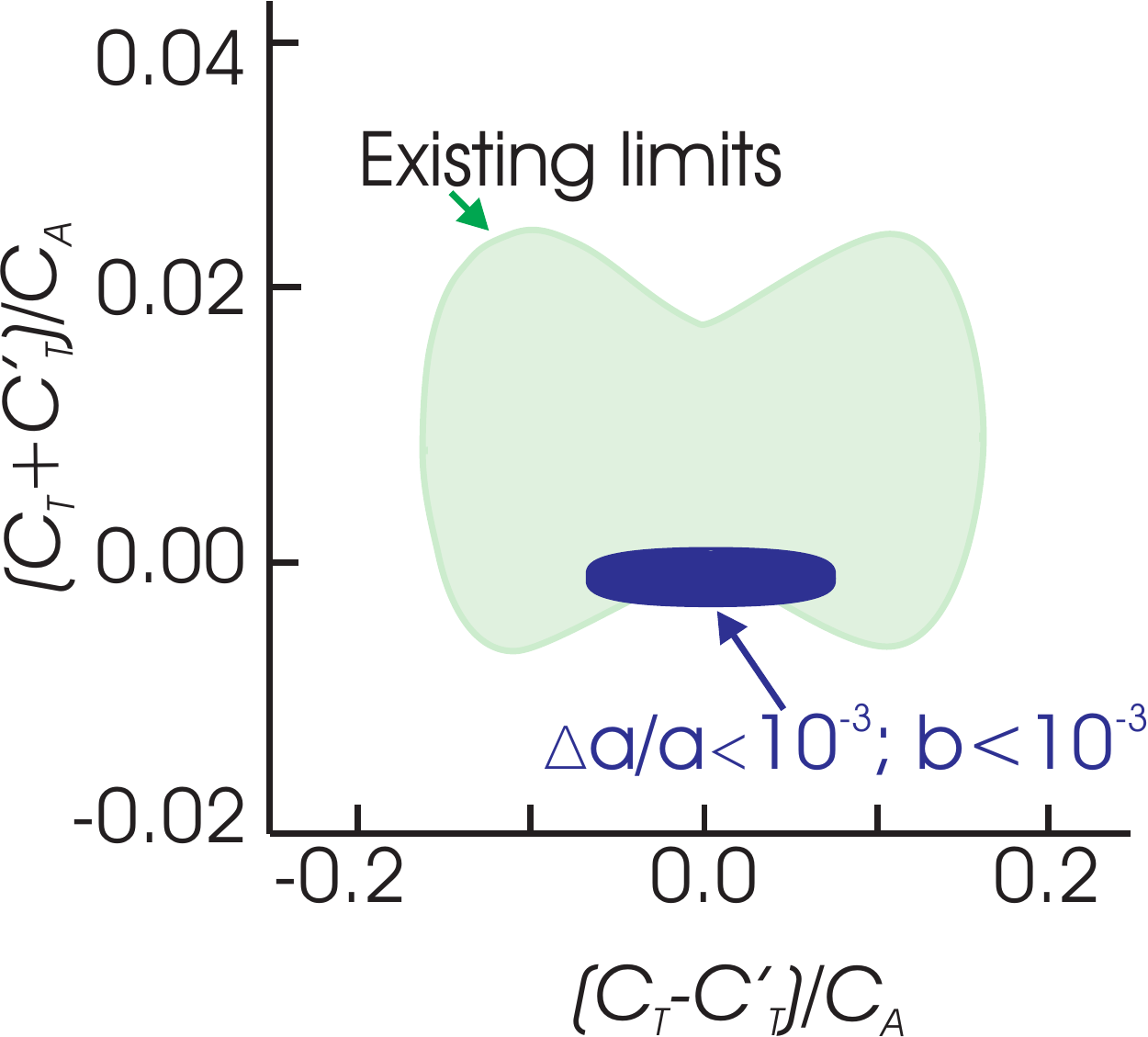}\hfil 
\centering
\vspace*{.5\baselineskip}
\caption{95\% confidence intervals: presently allowed region (from Ref.~\cite{se:06}) in green and,   in blue, the limits one would get with measurements of \protect $\Delta a/a < 10^{-3}$ and $\Delta b < 10^{-3}$.}
\label{fig: He6-limits}
\end{figure} 

In the $^{6}{\rm He}$ case, with a $0^+ \rightarrow 1^+$ beta decay, the connection between $a,\;b$ and Tensor currents is {\em direct}, as opposed to the case of mirror transitions, such as neutron decay, where $a$ and $b$ are sensitive in addition to the ratio of axial to vector current couplings and matrix elements plus potentially existing scalar currents. The decay of $^6{\rm He}$ presents an excellent opportunity for these searches. The large $Q$-value allows for an exploration of the shape of the spectrum over a broad energy range. By comparison, neutron beta decay presents an endpoint smaller by a factor of about 3.5 and the potential problem of capture of neutrons in the environment generating additional backgrounds. Being the lightest radioactive nucleus with a pure GT decay, the recoilling nucleus shows high sensitivity to the momentum carried by the leptons, which is good for a determination of the correlation coefficient.

The CENPA system produces copious amounts of $^6{\rm He}$ \cite{kn:11}: more than $10^9$ atoms per second delivered to a low-background room. Part of the program consists of using laser traps \cite{mu:11} to determine the electron-anti-neutrino correlation by coincidence detection of the beta and $^6{\rm Li}$ recoil ion. 
The collaboration presently already has a working MOT trap with $^{6}{\rm He}$ atoms.
Another part of the program is to determine the shape of the beta spectrum for which one is not planning to use the laser-trap system. An ISOL-type FRIB could provide up to three orders of magnitude higher production rates allowing for further improvements in both the $e,{\overline \nu}$ correlation and for the spectrum-shape measurement, which could then be done with trapped atoms.

\subsection{$\beta$-Decay with Ion Traps}
\label{sec:trapsII}

%\begin{figure}\centering
%\begin{wrapfigure}[15]{r}{0.5\textwidth}\centering
%  \includegraphics[width=0.45\textwidth]{/home/dmelconian/tamu/CI-layout.eps}
%  \caption{Layout of the \tamutrap{} beamlines and coupling to the 
%    high-intensity upgrade to the Cyclotron Institute at Texas A\&M 
%    University.\label{fig1}}
%\end{wrapfigure}
%\end{figure}

Penning traps of ions are best known for the incredible precision with 
which they can measure masses: relative uncertainties of $\Delta M/M\simeq
10^{-11}$ have been reported on stable species~\cite{pritchardNature}, and 
$\simeq10^{-8}$ for very short-lived ($\gtrsim 10$~ms) exotic 
ions~\cite{blaumPRL}.  These mass measurements have impacts in many fields 
of science, including fundamental physics research (CKM unitarity, testing 
nuclear models, correlation studies, etc.).  Penning traps are also used 
in other applications, including laser spectroscopy, QED effects, 
electron-capture studies and the astrophysical r-process, to name a few.  

The  group at Texas A\&M University are in the process of constructing a new 
double-Penning trap facility, \tamutrap, which will take advantage of the 
radioactive ion beam capabilities of the upgraded Cyclotron Institute 
facility, \trex~\cite{cyclotron-upgrade}.   The 
components that are currently being built as part of the \trex{} upgrade are: 
re-commisioning the K150 cyclotron to deliver high intensity light particle 
and heavy ion beams; the light and heavy ion guide systems; the charge-breeding 
ECR ion source and coupling of it to the K500 cyclotron, to provide high 
quality re-accelerated rare beams of both neutron and proton rich isotopes 
in the 5 to 50~MeV/u range.  

In addition to the superallowed program, 
the system is being designed to be flexible and allow for other fundamental 
physics studies, including the ability to perform precision mass measurements.  
Although not a user facility, the Cyclotron Institute has a long history 
of collaborating with outside users and we expect that with our existing and 
planned upgrade for extending our RIB capabilities, that collaborative efforts 
will continue to be formed with groups interested in using our facilities.

{\bf The Intensity Frontier:}\ \ 
Both atom and ion traps provide extremely clean sources, so experiments can 
be performed with relatively small sample sizes owing to the very large 
signal-to-noise ratio.  However, high intensities with relatively long access 
times to radioactive beams would be extremely advantageous to these 
experiments:  most are systematics-limited, where the systematics 
are themselves statistics limited; \emph{i.e.}\ systematic uncertainties could 
be considerably reduced if their sources could be investigated and minimized by 
using dedicated beamtime to characterize and quantify them.  The \trex{} 
facility at Texas A\&M University is not a user facility and so can offer 
better availability of beamtimes than, for example, F{\small RIB} or \triumf; 
however, the larger facilities can provide greater intensities.  Ideally, one 
would like to have a facility that can provide both high intensities and 
long accessibility of beamtimes.  Low-energy nuclear physics programs using 
both MOTs and Penning traps would be able to capitalize on such a facility and 
meaningfully add to probes of physics beyond the standard model.

%%%%%%%%%%%%%%%%8Li ion trap at ANL%%%%%%%%%%%%%%%%%
\subsection{Search for Tensor and 2nd-Class Currents in mass-8 systems}
\label{sec:mass8}

A search for tensor and second class currents is ongoing at ANL from a measurement of the beta-neutrino correlation in the decay of trapped radioactive ions stored in an open geometry ion trap. Low-energy neutrinos are extremely difficult to observe and in practice the beta-neutrino correlation is obtained via a kinematically complete measurement of the beta-recoil correlation.  An ideal source for such correlation measurements is a sample of radioactive ions at rest in high vacuum in a trap surrounded by a detector array. The Beta Paul Trap (BPT) is an ion trap system currently built for such measurements and currently being used at ANL to study the decay of $^8$Li and $^8$B. These decays feed an excited state in $^8$Be which breaks up into two alpha particles and the measurement of the 3 charged particles (2 $\alpha$’s and 1 $\beta$) allows a reconstruction of the neutrino momentum in an event by event basis, providing great sensitivity to the searched for currents. The power of the technique was demonstrated by a test experiment on the $^8$Li decay which yielded the second best low-energy limit on tensor currents~\cite{Li:2013}. The main sources of systematic errors identified in the test experiment were corrected and a production experiment was completed last year with 20 times more statistics~\cite{Sternberg:2013} and will yield the most sensitive tensor current low-energy limit. The universal injection system developed to feed radioactivity into the ion trap allows similar measurements to be performed on any species and a similar accuracy measurement is ongoing for the decay of $^8$B. In addition to a similar limit on tensor current, this experiment will also yield improved limits on second class currents and an improved test of the conserved vector current when combined to the $^8$Li results, and an improved $^8$B neutrino spectrum important to resolve a recent controversy which has important implications for solar neutrino physics.

\section{Neutral Currents}
\label{sec:Neutral}

Experiments using intense beams of polarized electrons scattering from fixed targets as well as those exploiting parity-forbidden atomic transitions have played a vital role in developing and testing the electroweak sector of the Standard Model as well as in probing novel aspects of hadron and nuclear structure (for a reviews, see Refs.~\cite{Cirigliano:2013lpa, Kumar:2013yoa, Erler:2013xha}).
Most recently, a program of parity-violating electron scattering experiments at MIT-Bates, Mainz, SLAC and Jefferson Laboratory have yielded stringent limits on contributions from the strange quark sea to the nucleon's electromagnetic properties and have provided the most stringent test to date of the running of the weak mixing angle $\theta_W$ below the weak scale. Similarly powerful measurements have been carried out with beams of atomic cesium, yielding the most precise determination of the nuclear weak charge and intriguing indications of the so-called nuclear \lq\lq anapole moment". 

These achievements have built on the steady improvements in experimental sensitivity since the pioneering measurements of parity-violating, deep inelastic electron-deuteron scattering~\cite{SMB} and atomic parity-violation in the 1970's, together with refinements of the theoretical interpretation. The frontier of this field now promises a unique capability to probe possible new physics at the TeV scale as well as previously inaccessible features of nucleon and nuclear structure. At the same time, the possibility to carry out a sensitive search for charged lepton flavor violation with unpolarized beams at an electron-ion collider (EIC) appears increasingly feasible. Below, we review some of the present efforts and future prospects for studies that exploit neutral current interactions with  electrons. 

To set the stage, we introduce the low-energy effective Lagrangian characterizing the parity-violating neutral weak electron-quark interaction:
\begin{equation}
\label{eq:leffpv}
\mathcal{L}^{PV}=\frac{G_F}{\sqrt{2}}
[\overline e\gamma^\mu\gamma_5e(C_{1u}\overline u\gamma_\mu u+C_{1d}\overline
  d\gamma_\mu d)
+\overline e\gamma^\mu e(C_{2u}\overline u\gamma_\mu\gamma_5 u+C_{2d}\overline
  d\gamma_\mu\gamma_5 d)].
\end{equation}
In the Standard Model, the coefficients $C_{1q}$ and $C_{2q}$ can be predicted with high precision and depend critically on 
$\sin^2\theta_W$. A similar expression is obtained for the parity-violating electron-electron interaction. One goal of the parity-violation experiments is to test these predictions. As with the muon anomalous magnetic moment, any deviation from the Standard Model predictions  would indicate the presence of new physics to the extent that the theoretical predictions are sufficiently robust. Given the present level of experimental and theoretical uncertainties, the present generation of PV experiments are able to probe for new TeV scale physics. 

From a complementary perspective, one may assume the Standard Model values for the $C_{iq}$ are correct and use the interactions in Eq.~(\ref{eq:leffpv}) to access properties of the nucleon and nuclei that are not readily probed by the the purely electromagnetic interaction. Indeed, the determination of strange quark contributions relies on different combinations of the light quark currents that enter electromagnetic and neutral weak currents. As discussed below, a number of additional interesting aspects of nucleon and nuclear structure can be uncovered using the parity-violating electron-quark interaction. 

\subsection{Proton Weak Charge}
\label{sec:Qweak}

The Qweak collaboration~\cite{Armstrong:2012ps} is conducting the first precision measurement of the weak charge of the proton, $Q^{p}_W$, given in terms of the $C_{1q}$ as
\begin{equation}
Q_W^p = -2\left(2C_{1u}+C_{1d}\right).
\end{equation} 
At leading order in the Standard Model, $Q_W^p=1-4\sin^2\theta_W$. 
This experiment was performed at Jefferson Laboratory, building on the technical advances made in the laboratory's parity-violation program and using the results of earlier measurements to constrain hadronic corrections. The experiment measures the parity-violating longitudinal analyzing power in e-p elastic scattering at $Q^{2}$ = 0.026 $(GeV/c)^{2}$ employing 180~$\mu A$'s of 86\%  polarized electrons on a 0.35~m long liquid hydrogen target. The measurement will determine the weak charge of the proton with about 4.1\% combined statistical and systematic errors. This corresponds to constraints on parity violating new physics at a mass scale of 2.3 TeV at the 95\% confidence level. This also allows $\sin^{2}\theta_{W}$ to be determined to 0.3\% accuracy, providing a competitive measurement of the running of the mixing angle.  

The parity-violating asymmetry,  $A_{PV}=(\sigma_R-\sigma_L)/(\sigma_R+\sigma_L)$
%$A_{LR}(^{1}H)$ 
 is the asymmetry in the measurement of the cross section difference between elastic scattering by longitudinally polarized electrons with positive and negative helicity from unpolarized protons. $Q^2$ is the four-momentum transfer, $\tau = Q^{2}/4M^{2}$ where $M$ is the proton mass,  and $\theta$ is the laboratory electron scattering angle. For forward-angle scattering where $\theta \rightarrow 0$, $\epsilon \rightarrow 1$, and $\tau << 1$, the asymmetry can be written as:
\begin{equation}
A_{PV} = 
\left[- G_F Q^2 \over 4 \pi \alpha \sqrt{2}\right] 
\left[  Q_{W}^{p} + F^{p}(Q^{2},\theta,E) \right] 
\rightarrow 
\left[- G_F Q^2 \over 4 \pi \alpha \sqrt{2}\right] 
\left[   Q_{W}^{p} + Q^{2} B(Q^{2})+C(E)\right]
 \label{arc22}     
\end{equation}
where $F^{p}$ is a form factor that includes a dependence on the beam energy $E$ beyond the Born approximation. The first term, proportional to $Q^2$, is for a point-like proton. The second term $B(Q^{2})$, proportional to $Q^4$, is the leading term in the nucleon structure defined in terms of neutron and proton electromagnetic and weak form factors. Ideally we would like to measure at a low enough $Q^2$ that the proton would look like a point and hadronic corrections would be negligible.  An accurate measurement of $\sin^{2}\theta_{W}${} thus requires higher order, yet significant, corrections for nucleon structure. Nucleon structure contributions in $B(Q^{2})$ can be suppressed by going to lower momentum transfer and energy. The numerical value of $B(Q^{2})$ has been constrained experimentally by extrapolation from existing forward angle parity-violating data at higher $Q^{2}$. The importance of the additional $E$-dependent  contributions arising from the exchange of two vector bosons between the electron and proton remains a topic of on-going theoretical scrutiny. 

\begin{figure}[ht]
\begin{center}
\rotatebox{0.}{\resizebox{3.7in}{3.3in}{\includegraphics{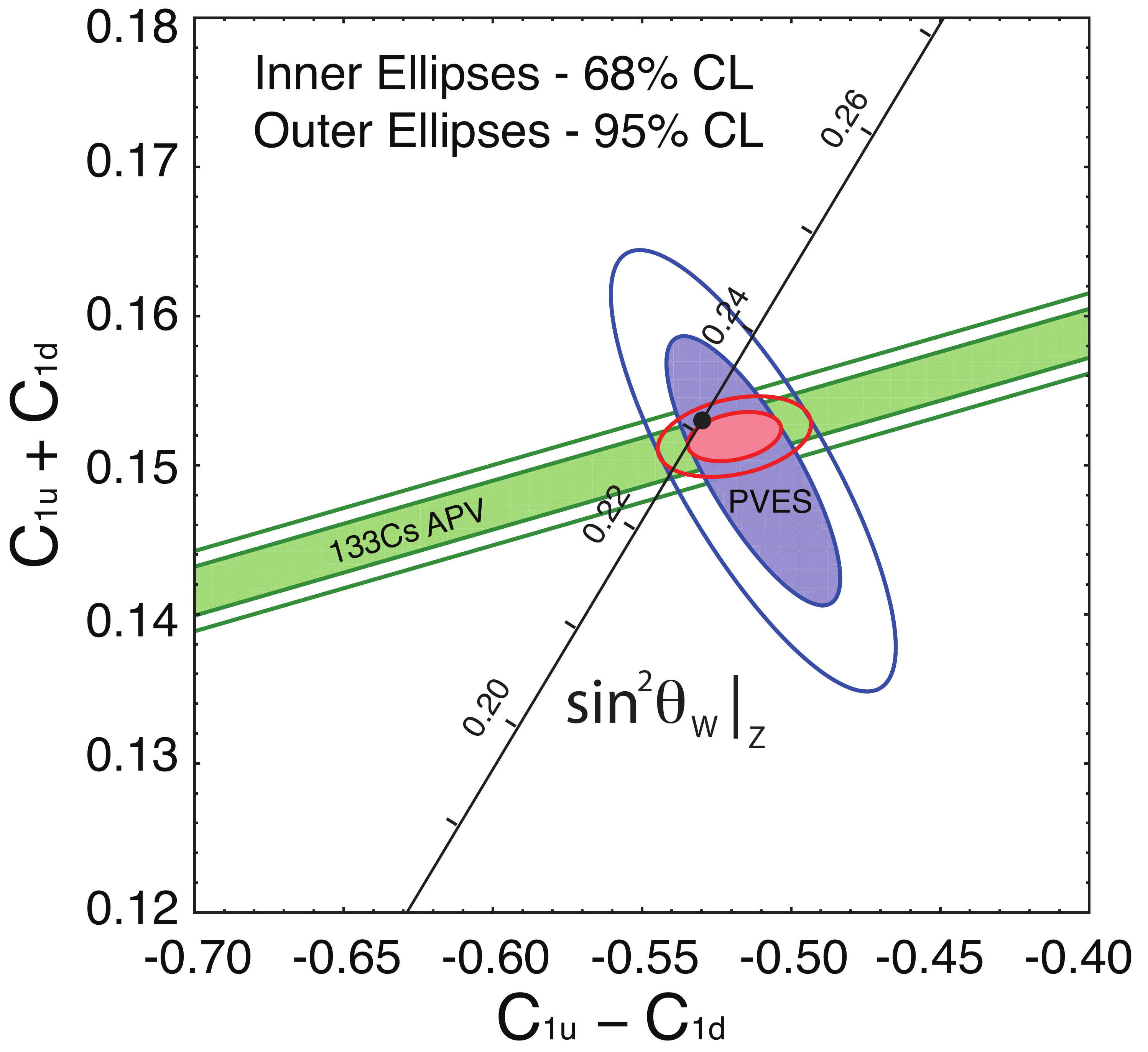}}}
\end{center}
\caption{ The constraints on the neutral-weak quark coupling constants {$C_{1u}- C_{1d}$} (isovector) and 
{$C_{1u} + C_{1d}$} (isoscalar). The more horizontal (green) APV band provides a tight constraint on the isoscalar combination from $^{133}$Cs data. The more vertical (blue) ellipse represents the global fit of the existing $Q^2 < 0.63$ PVES data including the new result reported here at $Q^{2}$=0.025 $({\rm GeV/c})^{2}$. The smaller (red) ellipse near the center of the figure shows the result obtained by combining the APV and PVES information. The SM 
prediction~\cite{Beringer:1900zz} as a function of $\sin^{2}\theta_W $ in the $\overline{MS}$
scheme is plotted (diagonal black line) with the SM best fit value indicated by the (black) point at $\sin^{2}\theta_W $=0.23116. (Reproduced from Ref.~\cite{Androic:2013rhu}) }
\label{searches-3}
\end{figure}

Analysis of the data is ongoing and the collaboration recently released the first measurement of the weak
charge using data from the commissioning run, comprising 4\%\ of the full dataset~\cite{Androic:2013rhu}. 
In combination with other parity-violation measurements, a high precision determination of the weak couplings $C_{1u}$ and $C_{1d}$, improving significantly on the present knowledge as shown in Figure~\ref{searches-3}. 

\subsubsection{Proposals for Improved Measurements}

Initial simulations indicate that it is technically feasible to perform  $\sim$2\% ultra low $Q^2$ measurement of $Q^{p}_W$ by using the existing apparatus with approximately 0.5 mA of 200 MeV beam potentially available at the JLab Free Electron Laser (FEL) accelerator complex. The value of such a measurement would really come from its impact on a global fit. It would be another point, at a much lower $Q^2$, but with different systematic and theoretical uncertainties. Such a measurement would have significantly smaller  "dilution" terms (magnetic moment, plus any strange quark, etc.) and considerably smaller $E$-dependent hadronic effects. 

Realizing this capability would require the construction of new endstation (similar in size to the existing Hall B at JLab), the addition of a polarized injector, beam polarimetry and some upgrade work on the FEL accelerator.  If the existing Qweak apparatus were simply used with 200 MeV beam it would have the following characteristics: $Q^2$ = 7 x $10^{-4}$ $(GeV/c)^2$, with an effective detector rate of 1.5~GHz/$\mu$A of beam incident on the hydrogen target.

The P2 experiment has been proposed for the newly funded MESA facility at Mainz to pursue a similar
measurement. The goal is $\delta(A_{PV})= \pm 1.7\%$ (stat. + syst.) for elastic electron-proton, which would yield 
$\delta(Q_W^p)\simeq 2\%$ and $\delta(\sin^2\theta_W)\pm 0.15\%$. 
To achieve the statistics would require
a 200 MeV, 150 $\mu$A beam incident on a 60 cm LH$_2$ target for 10,000 hours.
Such a measurement is even more ambitious than the JLab FEL concept described in the previous subsection
and has all the of the same advantages.  

The design requires a solenoidal magnet (such as the inner tracking field of the ZEUS collider detector at 
DESY) downstream of the target which would focus scattered electrons within 
$10^\circ<\theta_{\rm lab}<30^\circ$ in the full range of the azimuth onto integrating Cherenkov
detectors.  The field would sweep out the large M\o ller electron background and allow judiciously placed 
annular slits to shield the detectors from the target's direct photon background. 

The design must overcome many technical challenges such as controlling electron beam fluctuations at the
sub-nm level and controlling target density
fluctuations to a few parts in $10^{-5}$. A new method to measure the electron beam 
polarization must be developed, which would require a novel polarized hydrogen 
gas target.
The design and required R\&D will be carried out in the next few years so that the experiment
would be ready to start commissioning when MESA first produces external beams, anticipated for 2017.

%%%%%%%%%%%%% PV DIS %%%%%%%%%%%%

\subsection{Parity Violating Deep Inelastic Scattering}
\label{sec:pvdis}

%\section{Introduction}

At JLab, experiments are ongoing to measure parity violation in the deep
inelastic scattering (PVDIS) of polarized electrons from deuterium. 
The goals of the measurements are:
%\begin{enumerate}
(1) Measure the $C_{2q}$ coefficients with high precision;
(2) Search for charge symmetry violation (CSV) at the quark level;
(3) Search for quark-quark correlations in the nucleon.
%\end{enumerate}

\subsubsection{Physics of PVDIS}

In 1978, Prescott et al.~\cite{SMB} showed that parity-violating could be
observed in neutral currents by measuring the asymmetry
%\begin{equation}
$A_{PV}=(\sigma_R-\sigma_L)(\sigma_R+\sigma_L)$
%\label{eq:apvdef}
%\end{equation}
in the deep inelastic scattering (DIS) of polarized electrons from deuterium.
By extending the kinematic range, the same group later published results that
were able to exclude alternative theories to the Standard Model that were
considered reasonable at the time.  The PV DIS asymmetry 
is sensitive to both the axial-vector (vector) electron couplings 
$C_{1q}$ and $C_{2q}$, whereas the emphasis for the Q-Weak experiment is on the $C_{1q}$.
%quarks in the phenomenological Lagrangian  
For deuterium, one has
\begin{equation}
A^{PV}=
-\left(\frac{G_FQ^2}{4\sqrt{2}\pi\alpha}\right)
\left[a_1+\frac{1-(1-y)^2}{1+(1-y)^2}a_3\right];\ \ 
a^D_1(x)=
-\frac{6}{5}(2C_{1u}-C_{1d});\ \ 
a^D_3(x)
=-\frac{6}{5}(2C_{2u}-C_{2d}),
\label{eq:Afull}
\end{equation}
where $y=\nu/E$.  By observing the dependence of $A_{PV}$ on $y$, the Prescott
experiment was sensitive to both the $C_{1i}$'s and the $C_{2i}$'s.

As discussed above, there has been great progress more recently in the precision of the
measurements of the $C_{1q}$'s with the advent of the Qweak experiment and
precision atomic physics measurements.  On the other hand, progress with the $C_{2i}$'s has been
slower because at low energies, uncertain radiative corrections involving
the long-distance behavior of hadrons are large.
However, in DIS, these corrections are tractable since the relevant energy scale implies that all corrections can be computed perturbatively.  Hence the motivation for a
new precise experiment in PVDIS from deuterium.  The sensitivity of
the JLab experiment to new physics is given by 
Kurylov et al.~\cite{Kurylov:2003xa}.

\subsubsection{An Improved Measurement of PVDIS using the JLab 6 GeV Beam}

The first measurement of $A_{PV}$ in deep-inelastic scattering since the original SLAC E122 measurement 
discussed previously was carried out by JLab experiment 
E08011. 
The primary motivation was to constrain the poorly known $C_{2i}$
couplings. 
The experiment ran in late 2009 with an incident beam energy of 
$\sim 6$ GeV and $Q^2\sim 1-2$ GeV$^2$, collecting sufficient statistics to measure $A_{PV}$ off $^2$H with a 
fractional accuracy better than 4\%. 

The scattered electrons were detected by the Hall A High Resolution Spectrometer (HRS) 
pair. Unlike other high rate experiments discussed in this review, a custom fast counting
data acquisition system was used. Event-by-event particle identification was carried out at the hardware level 
with gas Cherenkov detectors and lead-glass shower counters. This information was fed into fast trigger logic
to form electron and pion triggers that were in turn fed into scalers. The electron scaler results over the duration
of each helicity time window of the electron beam were used to construct the raw asymmetry from which $A_{PV}$
could be extracted. The electron trigger efficiency was found to be greater than 95\%, with a pion rejection 
$> 1000:1$. Data analysis is ongoing and final results are expected to be published by late 2013.

\subsubsection{SOLID: A Proposal using a 11 GeV Beam at the Upgraded JLab Facility}

\begin{figure} [htbp]
\begin{center}
\rotatebox{0.}{\resizebox{6.0in}{2.5in}{\includegraphics{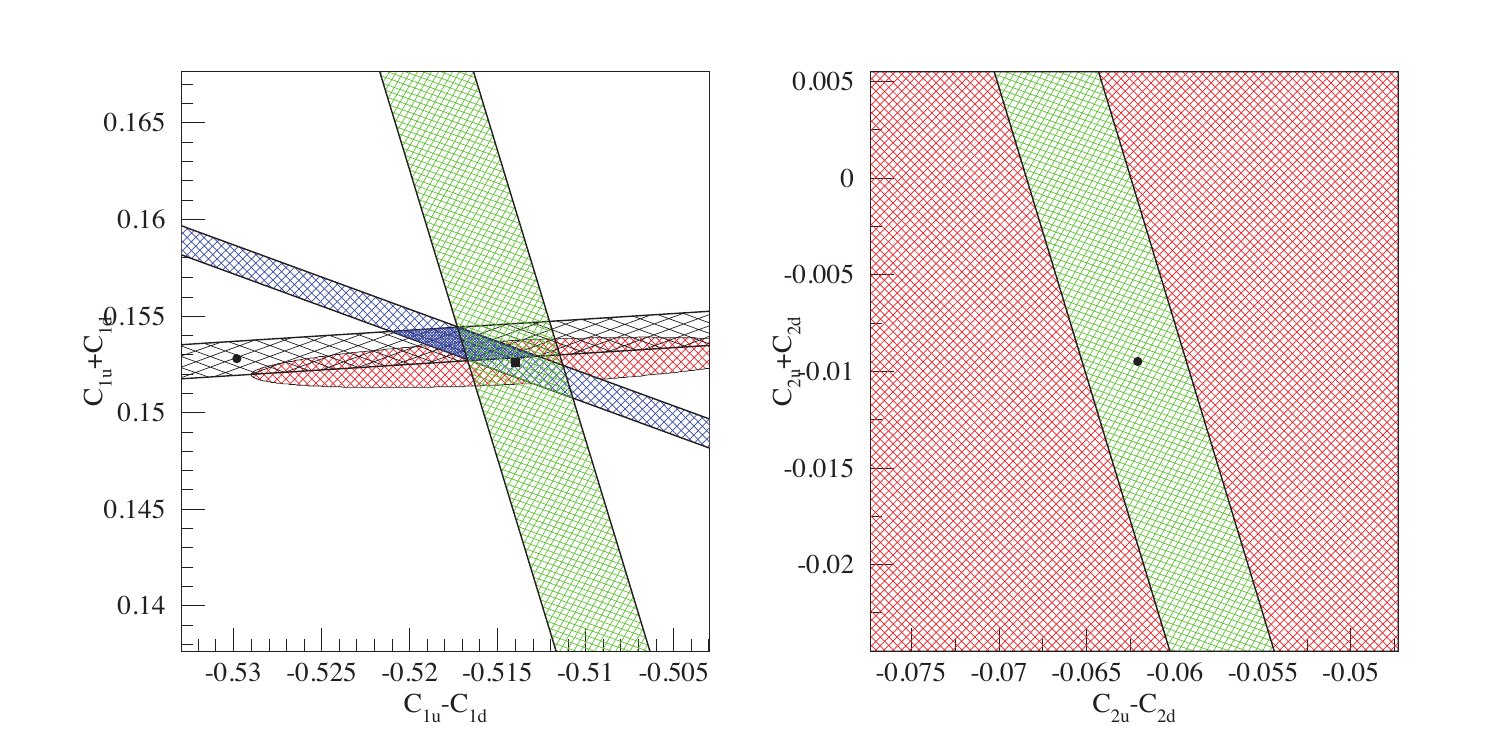}}}
\end{center}
\caption{ Left plot: Constraints on the Standard Model from
         parity-violation experiments.
         The black crossed band presents the
         results from atomic parity violation in Cs atoms, the blue band 
         is the expected result from Qweak, the red ellipse
         is a PDG fit, 
         while the green band shows the anticipated JLab limit.
         Right plot: The anticipated error band from the JLab experiment.
         All limits are 1 standard deviation.}
\label{fig:pvdis}
\end{figure}

At JLab, there are plans  to build a new solenoidal spectrometer called SoLID that
will enable one to obtain statical precision of $<0.5$\% for a number of bins 
with $ 0.3<x<0.75$, $4<Q^2<10$ (GeV/c)$^2$, and $y\sim 1$~\cite{SOLID}.  
By designing the
apparatus for large $y$, one can maximize our sensitivity to the $C_{2i}$'s.
The  expected sensitivity  is given in Figure~\ref{fig:pvdis}.

Presently the CLEOII magnet is being
considered for the solenoid.  Polarized electrons with an energy of 11 GeV will strike a liquid
deuterium target.  Scattered lectrons with energies above 2 GeV will pass 
throurgh a series of baffles and
then strike a detector package consisting of tracking chambers, a Cerenkov
counter to reject pions, and a calorimeter to serve as a trigger and provide
additional pion rejection.  The data will be taken with a deadtimeless flash 
ADC system with 30 independent sectors.  The tracking detectors will be
GEM's with a total area of 25 m$^2$~\cite{Ketzer:2004jk}.

\subsubsection{Hadronic Corrections}

The NuTeV experiment published a test of the Standard Model in neutrino-nucleus DIS .  The result differed from the prediction of the Standard Model
of about $3\sigma$~\cite{Zeller:2001hh},
generating considerable controversy and a causing a serious re-evaluation of the 
work.  Corrections, including
changes in the value of $V_\mathrm{us}$, strange sea, and improved radiative corrections
have been made.  Including these effects has not substantially mitigated the discrepancy.
Another possible explanation of the NuTeV result is charge symmetry violation 
(CSV) in the parton distribution functions (PDFs)\cite{Londergan:2003pq,Londergan:2003ij}.
Various authors have presented the case that this is a reasonable
explanation, citing the effects of PDFs \cite{Martin:2003sk}, QCD 
effects \cite{Sather:1991je,Rodionov:1994cg,Cloet:2009qs,Bentz:2009yy},
and QED effects \cite{Martin:2004dh,Gluck:2005xh}.

The JLab PV DIS  experiment is also sensitive to CSV.  If the $x$ dependence of the
CSV falls more slowly than that of the PDFs, the asymmetry should display a clear
$x$ dependence.  Moreover, these results will provide an important
test of the CSV explanation for NuTeV.  More details are given in the 
proposal~\cite{SOLID}.

%\begin{figure}[htb]
%  \includegraphics[width=0.7\columnwidth]{C2PVDIS.pdf}
%  \includegraphics[width=0.45\columnwidth]{c2pm_las_small.eps}
%  \caption{Left plot: Constraints on the Standard Model from
%         parity-violation experiments.
%      The black crossed band presents the
 %    results from atomic parity violation in Cs atoms, the blue band 
 %  is the expected result from Qweak, the red ellipse
 %  is a PDG fit, 
  % while the green band shows the anticipated JLab limit.
  % Right plot: The anticipated error band from the JLab experiment.
  % All limits are 1 standard deviation.
  % }
%  \label{fig:SM_small}
% \end{figure}

There are additional important corrections to 
Equation~\ref{eq:Afull}~\cite{Hobbs:2008mm}. 
In particular, the cross sections
in this kinematic angle have significant scaling violations due to
higher-twist effects~\cite{Blumlein:2008kz}.  
However, as pointed out by Bjorken~\cite{Bj} and 
more recently by Mantry et al.~\cite{Mantry:2010ki}, the higher twist 
terms cancel in the $a_1$ term unless they are
due to quark-quark correlations.  The observation of diquarks in the nucleon,
if found, would be an exciting discovery.  The higher twist contribution of the
$a_3$ term can be subtracted from our asymmetry by using data on charged 
neutrino scattering~\cite{Bj}. 

%%%%%%%%%%%%%%%% PV Moller %%%%%%%%%%%%%%%

\subsection{Parity Violating M\o ller Scattering}
\label{sec:moller}

An observable that has the potential to search for new flavor conserving amplitude as small as
$10^{-3}\times G_F$ in a purely leptonic process is 
$A_{PV}$ in electron-electron (M\o ller) scattering.  
The SLAC E158 experiment made the first ever measurement of this observable~\cite{e158}, 
and set
important complementary limits on contact interactions with a sensitivity comparable to the highest energy
colliders that were then running in parallel i.e. LEP-II and the Tevatron. 

Recently, the MOLLER experiment has been proposed at Jefferson Laboratory to improve on the E158
measurement by more than a factor of 5. The goal is a measurement of $A_{PV}$ and thus 
the weak charge of the electron
$Q^e_W$, which is proportional to the product of the electron's vector and axial-vector couplings to the $Z^0$ boson, to a fractional accuracy of 2.3\%. The prediction for $A_{PV}$ for the proposed experimental design is 
$\approx 35$~parts per billion (ppb) and our goal is to measure this quantity with a statistical precision of 0.73 ppb
and thus achieve a 2.3\%\ measurement of $Q^e_W$. The reduction in the numerical value of $Q^e_W$ 
due to radiative corrections leads to increased fractional accuracy in the determination of the weak mixing
 angle, $\sim 0.1$\%, comparable to the two best such determinations from measurements of asymmetries in
$Z^0$ decays in the $\mathrm{e}^+\mathrm{e}^-$ colliders LEP and SLC. 

\begin{figure}[ht]
\begin{minipage}[b]{0.48\linewidth}
\centering
\includegraphics[width=0.95\linewidth]{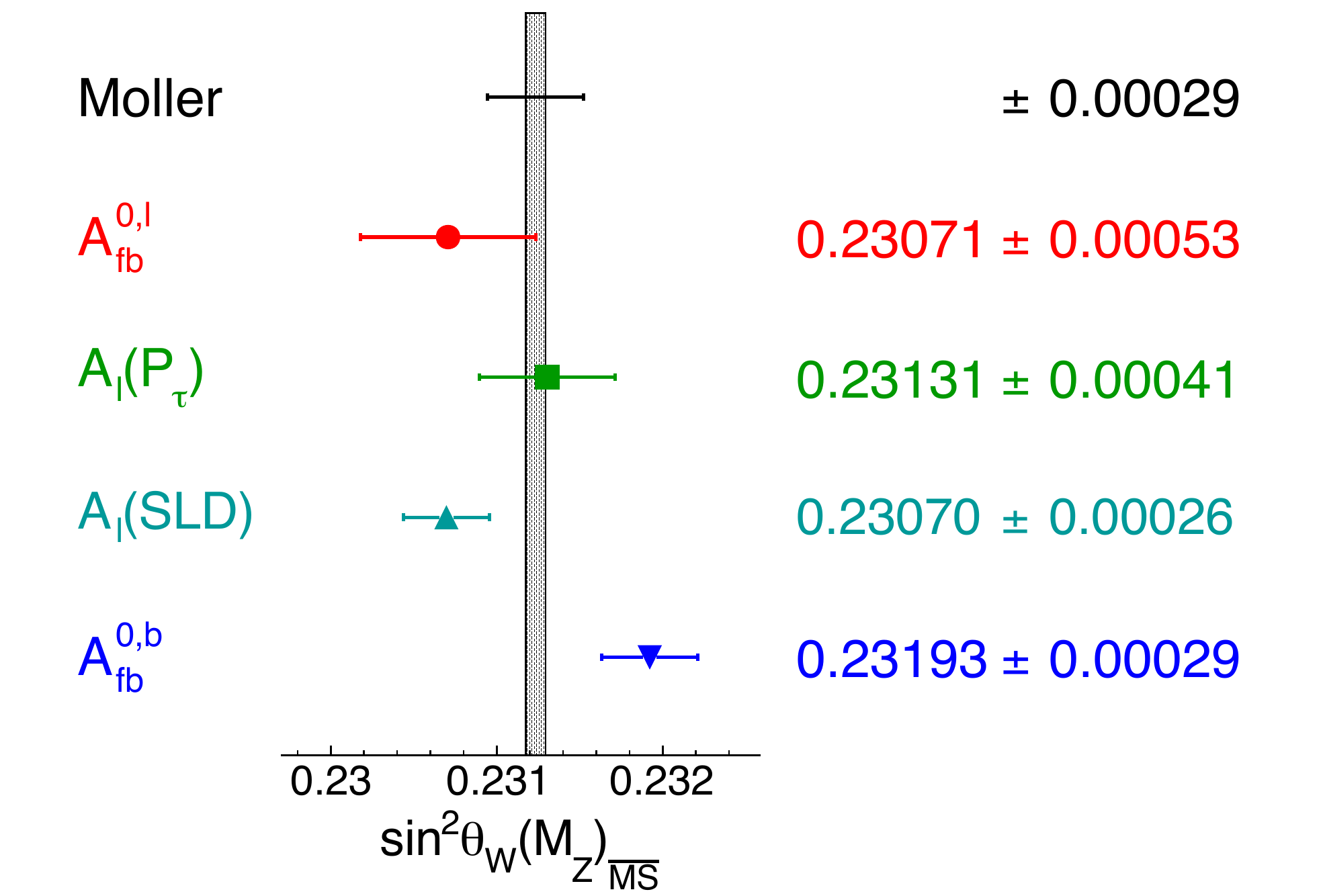}
  \caption{{\it The four best $\sin^2\theta_W$ measurements and the projected error of the MOLLER proposal.
  The black band represents the theoretical prediction for $m_H = 126$ GeV.}}
\label{fig:cl:s2tw}
\end{minipage}
\hspace{0.3cm}
\begin{minipage}[b]{0.48\linewidth}
\centering
    \includegraphics[width=3in]{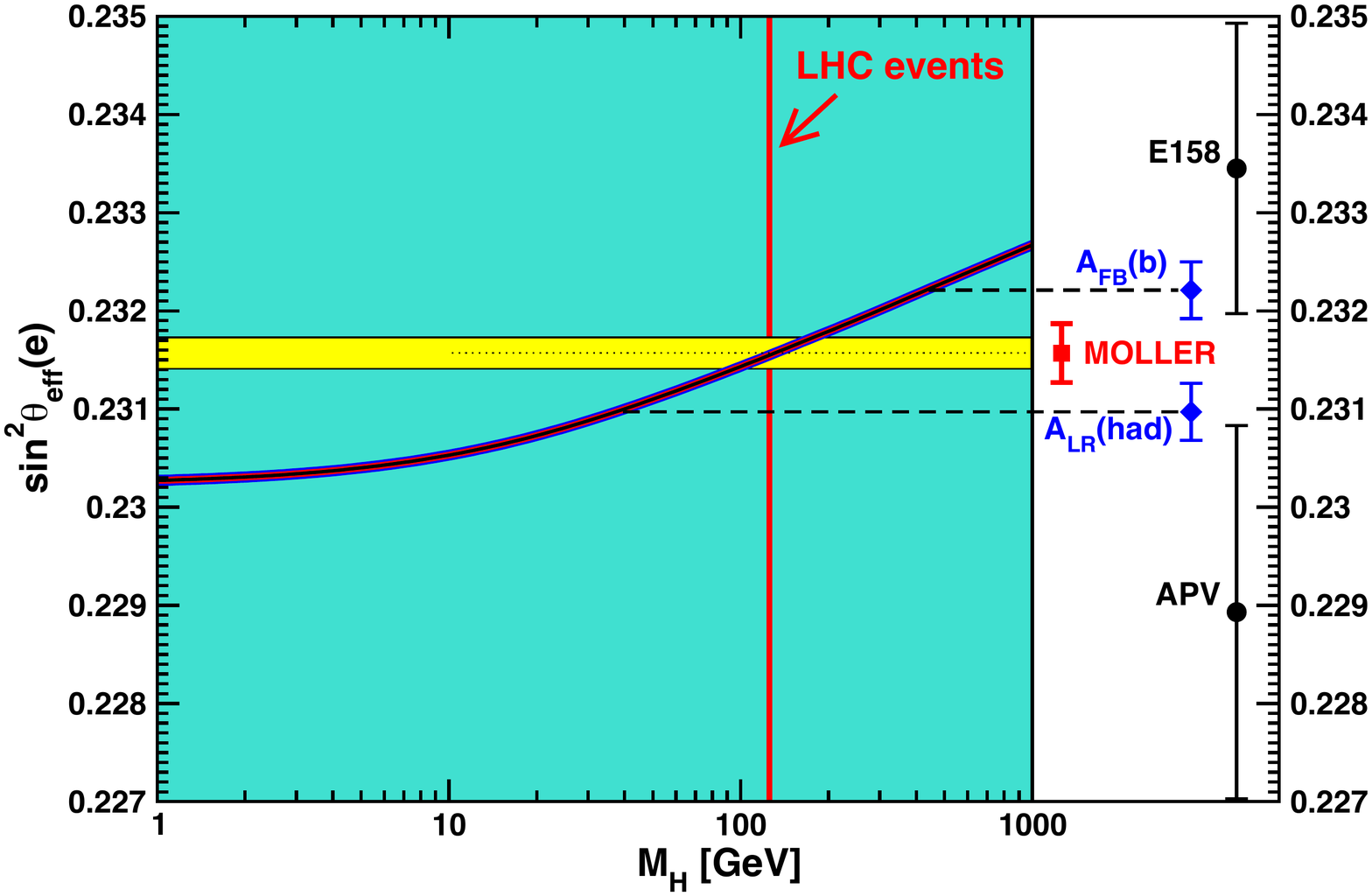}
  \caption{{\it $\sin^2\theta_W$ vs $m_H$. The yellow band shows the world average. The blue data points
  represent the two best high energy determinations while the black points are the most precise low energy
  determinations. The projected MOLLER error is shown in red. Figure courtesy of J. Erler.}}
  \label{fig:cl:s2twvsmh}
\end{minipage}
\end{figure}

Figure.~\ref{fig:cl:s2tw} shows the four best measurements from studies of $Z^{0}$ 
decays~\cite{ALEPH:2010aa}
as well as the projected uncertainty of the MOLLER proposal. Also shown
is the Standard Model prediction for a Higgs mass ($m_H$) of 126 GeV. It can be seen that the grand average of the
four measurements is consistent with the theoretical expectation. However, the scatter in the measurements
is disconcerting; it would be very useful to have new measurements such as MOLLER with comparable precision. 
Indeed, it is going to be very difficult to achieve this sensitivity with any other method 
until the advent of new facilities, which are more than a decade away.

An additional important feature of the proposed measurement is that it will be carried out at $Q^2\ll M_Z^2$.
The two best measurements of the weak mixing angle at lower energies are those extracted from the
aforementioned SLAC E158 measurement~\cite{e158}, 
and the measurement of the weak charge of $^{133}$Cs~\cite{wood97} via studies of table-top atomic parity violation. The interpretation of the latter measurement in terms of an extraction of the weak mixing angle has
been recently updated~\cite{Dzuba:2012kx}. Figure~\ref{fig:cl:s2twvsmh} shows the dependence 
of $\sin^2\theta_W$ to $m_H$ and the two best high energy and low energy measurements discussed above.
Also shown is the projected error for the MOLLER proposal. Remarkably, many different versions of new dynamics
at the TeV scale can have a significant impact on low $Q^2$ measurements while having little impact on
$Z^0$ decay measurements because interference effects on resonance are suppressed. It can be seen that there
is still plenty of room for new physics effects to be discovered at low energy given the proposed $A_{PV}$ 
uncertainty. 

At the level of sensitivity probed, the proposed measurement could be influenced by radiative loop effects of new 
particles predicted by the Minimal Supersymmetric Standard Model (MSSM). The impact on the weak charges of the
electron and the proton $Q_W^{e,p}$ have been analyzed in detail~\cite{Kurylov:2003zh}. A combined analysis of 
precision low energy measurements of both charged and neutral current processes can be found in a comprehensive review~\cite{RamseyMusolf:2006vr}, which has been recently updated~\cite{Erler:2013xha}. 
Inspecting a random scan over a set of MSSM parameters whose values are consistent with current precision 
measurements as well as the most recent LHC search limits from 7 and 8 TeV running, $A_{PV}$ would see 
in the effects in the range of 2 and 3 $\sigma$ at larger values of the MSSM parameter $\tan\beta$ (the ratio of vacuum expectation values of the model's two Higgs scalars) or if one of the superpartner 
masses is relatively light. 
If the assumption of R-parity conservation is relaxed (RPV), tree-level interactions could generate even deviations 
in $A_{PV}$ of opposite sign and similar magnitude. Thus, if nature is supersymmetric, the proposed measurement
would shed light on an important followup question regarding the validity of R-parity symmetry.

A comprehensive analysis of the MOLLER sensitivity to 
TeV-scale $Z^\prime$s has been carried out~\cite{Erler:2011iw} 
for a fairly large class of family-universal models
contained in the $E_6$ gauge group.  While models with full $E_6$ unification are already
excluded by existing precision electroweak data, the Z' bosons in these models with the same electroweak charges to SM particles are still motivated because 
they also arise in many superstring models as well as from a bottom-up approach~\cite{Erler:2000wu}. 
$A_{PV}$ probes
 $M_{Z^\prime} $ of order 2.5 TeV, comparable to the anticipated reach of 
early LHC running after the energy ramp-up to 13 TeV. The reach of $A_{PV}$ would be further 
enhanced in comparison 
to direct searches if one relaxes the model-dependent assumption of GUT coupling strength; indirect 
deviations scale linearly with other values of the coupling strength whereas dilepton production at colliders
has a much milder dependence on this parameter. 

%\section{Current Status|}
The 2007 NSAC long range plan report  comprehensively described the opportunities presented by new sensitive indirect probes such as MOLLER, and how they fit into the subfield of Fundamental Symmetries.
One of the overarching questions that serves to define this subfield is: ``What are the unseen forces that were present at the dawn of the universe but disappeared from view as the universe evolved?". To address this question and as part of the third principal recommendation, significant new investments, including MOLLER, were advocated.

MOLLER received the highest rating from the JLab Program Advisory Committee (PAC) in January 2009.
Subsequently in January 2010, 
JLab management organized a Director's review of the experiment chaired by Charles Prescott. The committee
gave strong endorsement to the experiment and encouraged the collaboration and the laboratory to develop
a full proposal to obtain construction funding.  In January 2011, 
the PAC allocated MOLLER's full beamtime request of 344 PAC days. 
More recently, the 2012 NSAC subpanel on the implementation of the Long Range Panel (the Tribble 
 Subcommittee) strongly endorsed the MOLLER project as part of the suite of investments advocated for 
 the subfield of Fundamental Symmetries.
JLab submitted a Major Item of Equipment (MIE) proposal
to DoE on behalf of the MOLLER collaboration ($\sim 100$ physicists from 30 institutions).
The goal is to obtain construction funding by 2015, with 
the hope of installing the apparatus and commissioning the experiment in 2017/18. 

\subsection{Atomic Parity Violation }
\label{sec:eic}

 Atomic parity-violation (APV) comes from the mixing of opposite parity states by the weak interaction that allows forbidden electro-magnetic (EM) transitions. The smallness of the weak interaction is enhanced experimentally by interference of an allowed and a forbidden transition, that then changes as the parity of the apparatus where it is measured changes. It has a rich history that spans more than thirty years \cite{khriplovich91,bouchiat00}, but continues to attract attention as its has been identified as a low energy area where it is possible to search for  `new' physics beyond the Standard Model
(SM) \cite{marciano90}. 

APV arises from the parity-violating exchange of $Z^{0}$-bosons between electrons and the
quarks in the nucleus \cite{bouchiat74}. The most precise measurement to date of APV was completed during the decade of the nineties in Boulder by the 
the group of  Wieman in Boulder in $^{133}$Cs \cite{wood97}.  
The  extraction of weak interaction physics requires a theoretical input that has been improving significantly during the last decade ({\it e.~g.} Ref.~\cite{ginges04,porsev09}).

APV measures the strength of the weak neutral current at 
very low momentum transfer. 
There are three types of such ``low-energy" weak neutral current measurements
with complementary sensitivity. The atomic weak charge is predominantly 
sensitive to the weak charge of the neutron, as the proton weak charge is 
proportional to $(1-4 \sin^2{\theta_W})$ which accidentally is near zero. 
The Qweak electron scattering experiment on hydrogen 
will be sensitive to the weak 
charge of the proton (see Ref.  \cite{blunden11} for a recent result on weak charge calculations). The SLAC E158 Moeller scattering is sensitive to the electron's weak 
charge. 
Different Standard Model extensions then contribute 
differently~\cite{ramsey-musolf08}. 
The atomic weak charge is relatively insensitive to one-loop order
corrections from all SUSY particles, so its measurement provides a benchmark
for possible departures by the other low-energy observables. 
Moeller scattering is purely leptonic and has no
sensitivity to leptoquarks, so APNC can then
provide the sensitivity to those.

{\bf Current ongoing efforts:} 
APV scales with the nuclear charge roughly as Z$^{3}$, favoring
experiments in heavy atoms (see the recent results in Yb in Ref.~\cite{tsigutkin09}). 
Ongoing efforts in APNC include, but are not limited to: Yb (Berkeley \cite{tsigutkin09}), Fr (FrPNC Collaboration at TRIUMF, and the Ferrara, Legnaro, Pisa, Siena collaboration), Ra$^+$ (KVI), and Dy (Berkeley). There is an overall plan to work with different isotopes, as a way to understand possible neutron distribution effects~\cite{brown09}, but also for studies of the weak interaction inside the nucleus, through anapole moment measurements (see for example \cite{gomez06} and references therein).

{\bf Exploration of new methods to measure APV:} 
All successful experiments that have measured APV have used an interference between an EM allowed transition and the weak, parity mixing, amplitude. The result comes then from the extraction of a rate of transitions with one or the other parities. This applies also to the experiments that measure optical rotation. The measurements are subject to many complications, that have been overcome. Many atomic physics and precision metrologists have developed great tools to measure frequency and frequency shifts. However, there is no frequency shift associated with a transition dipole. an electric dipole P-odd and T-even cannot give rise to a frequency shift in a stationary atomic state perturbed by homogeneous E and B dc fields. Important proposals have appeared and are beginning to be followed, that change the measurement of APNC from a rate to a frequency shift, with the use of quadrupole transitions in a single ion \cite{fortson93} or by using light shifts that can create a linear Stark shift measurable with matter-wave interferometry  \cite{bouchiat07,bouchiat08}.

{\bf Intensity frontier:} The successful APV experiments require as many atoms as possible ($N$). Current efforts with rare isotopes incorporate laser trapping and cooling to ensure large samples for interrogation \cite{gomez06}, ensuring the large $N$ regime. Methodologies for the measurement are exploring many new avenues, for example two-photon transitions \cite{duonas-frazer11}, and the proposals of Fortson and Bouchiat.  For radioactive ions, large numbers may be more complicated, given the effects of space charge on a cloud of ions; but care can be taken to pursue this avenue.The planned energy and current for Nuclear studies in Project X at Fermi Lab would allow to produce many orders of magnitude more rare isotopes than in any other facility in the world. If the design allows for parallel operation of multiple users. To achieve this will require development of appropriate targets and handling facilities. The flux, together with the multiuser parallel operation mode has the potential to become a fantastic tool for APNC research in the future.

\subsection{Hadronic Parity Violation}

While the focus of this document has been on the search for physics beyond the Standard Model (BSM), fundamental symmetry tests with nucleons, nuclei, and atoms have a rich tradition as probes of the strong interaction. One of the most interesting components of this program has been the study of the non-leptonic weak interaction of light quarks. In the strangeness changing ($\Delta S=1$) weak interaction, a number of puzzles have emerged over the years that still defy a satisfactory resolution: the origin of the $\Delta I=1/2$ rule; the incompatibility of the S- and P-wave amplitudes for non-leptonic hyperon decays; and the larger-than-expected breakdown of Hara's theorem in radiative hyperon decays. While BSM physics is undoubtedly not responsible for these puzzles, the culprit could lie with the presence of the strange quark, whose mass of order the QCD scale renders the application of either perturbative QCD or an effective field theory problematic at best. In this regard, it is interesting to study the non-leptonic weak interaction in systems where the strange quark plays a significantly less important role. In doing so, one can hope to learn whether the puzzles in the $\Delta S=1$ sector are unique to strange quark dynamics or reflect a more general aspect of the strong-electroweak interaction interplay.  

The study of parity-violating observables in nuclei and few-body systems provides such a window on the strangeness conserving non-leptonic weak interaction (for recent reviews, see Refs.~\cite{Haxton:2013aca,RamseyMusolf:2006dz}). At the lowest energies, the interaction is characterized by five S-P wave mixing amplitudes associated with the various isospin channels for the two-nucleon interaction. On such amplitude is particularly sensitive to long-range effects associated with pion exchange. The others are dominated by physics at the GeV scale and can be parameterized by effective NN contact interactions or by a heavy-meson ($\rho$ and $\omega$) exchange model. The most significant constraints on the parameters of these amplitudes have been obtained from elastic $pp$ scattering, elastic $p$-$\alpha$ scattering, PV transitions in light nuclei, and cesium APV that accesses the nuclear anapole moment. One puzzle that has emerged from this program is that the long-range pion exchange amplitude is significantly suppressed with respect to theoretical expectations. Whether or not this is the result of a nuclear renormalization or physics at the hadronic scale remains an open question. An ongoing experiment measuring the photon asymmetry $A_\gamma$ in ${\vec n}+p\to d+\gamma$ at the Spallation Neutron Source Fundamental Neutron Physics Beamline (FNPB) is designed to address this question. First results for the associated PV $\pi NN$ coupling have also been obtained using the lattice. A future, state-of-the art computation, combined with results of the $A_\gamma$ measurement, could clarify the picture of the long-range, $\Delta S=0$, non-leptonic weak interaction. More generally, a global analysis of hadronic PV results also reveals that presently phenomenology is essentially characterized by three combinations of the five independent S-P amplitudes\cite{Haxton:2013aca}. 

Looking to the future, it is desirable to complete a set of experiments that would over-determine the parameters of the low-energy S-P amplitudes. Comparing with lattice QCD computations of these parameters could then yield deeper insights into this long-standing problem. At the same time, a determination of the \lq\lq primordial" PV NN interaction could then be used to revisit the PV observables in nuclei. Doing so may provide a clearer window on the behavior four-quark operators in nuclei, potentially leading to better theoretical control over similar dynamics associated with heavy particle-mediated neutrinoless double $\beta$-decay. On the experimental side, a measurement of the rotation of neutron spin in $^4$He has been completed at NIST, though the present level of precision is not yet adequate to affect significant the global analysis. Additional possibilities include measurement of the photon circular polarization in unpolarized $np$ capture; $A_\gamma$ in ${\vec n}+t\to d+\gamma$; proton spin rotation in $^4$He; polarized ${\vec n}$ capture on $^3$He (approved for the FNPB); neutron spin rotation in deuterium; $n{\vec d}$ scattering with photons in the final state; and  deuteron photo disintegration with polarized photons. These possibilities are at various stages of development, and apart from the $n\, ^3\mathrm{He}$ experiment, none have proceeded to a formal proposal. We refer the reader to the recent NSAC neutron physics subcommittee report\cite{nsacneutron}, where additional details can be found. For purposes of this document, we note that one two-page white paper was received for a possible ${\vec\gamma}d\to np$ measurement at the upgraded HIGS (HIGS2) facility~\cite{higs2writeup}.
The High Intensity Gamma-Ray Sources (HIGS) is a Free-Electron Laser (FEL) based Compton backscattering facility. HIGS2 is a proposed upgrade that would provide unprecedented photon luminosity, polarization control and energy resolution. Together with future opportunities at the FNPB and NIST, a HIGS2 experiment has the potential to add significant input to the hadronic PV picture. At the same time, future work with lattice QCD to compute both the long- and short-range components of the PV NN interaction, in parallel with ongoing developments in few-nucleon effective field theory methods, is essential to the hadronic PV program.

\subsection{Electroweak Physics at an EIC}
\label{sec:eic}

Over the last decade, the Electron Ion Collider (EIC) has been considered in the US nuclear science community as a 
possible future experimental facility (beyond 12 GeV upgrade of the CEBAF at Jefferson Laboratory, and the FRIB at MSU) \cite{lrp07}
for the study of QCD. The EIC will help us explore and understand some of the most fundamental and universal aspects of
QCD \cite{lrp07, intreport}. 
%It will enable the most precise measurements yet of gluons and sea quarks, which are critical to our 
%understanding of constitution of the matter in the visible universe, including its fundamental properties such as mass 
%and spin.  If realized the physics program at the EIC will allow us to map out the 3D tomographic picture (position 
%correlated momentum distribution of partons) inside the nucleons and the nuclei. 
This physics program requires the EIC 
to have a variable center-of-mass energy from about $\sqrt{s}=30-140$ GeV, and the luminosities of 
$\sim 10^{33-34} {\rm cm}^{-2} {\rm sec}^{-1}$. 
for e-p collisions (100-1000 times that achieved at HERA), and polarization in both beams, and a wider range in 
nuclear species. The planned precision studies of QCD and the partonic dynamics also require the construction of 
a comprehensive detector system capable of excellent particle identification over a large momentum range, high 
momentum \& energy resolution and almost full (4$\pi$) acceptance. Currently there are two designs under consideration 
for the EIC in the US:
\begin{enumerate} 
\item  eRHIC \cite{eRHIC} at Brookhaven National Laboratory (BNL) which will use the hadron \& nuclear 
beams of the existing Relativistic Heavy Ion Collider (RHIC). The plan is to build an ERL based electron beam facility 
of variable energy 5-30 GeV in the existing RHIC tunnel to collide with one of the RHIC beams.
\item ELIC at Jefferson Laboratory (JLab) \cite{elic} 
which will use the electron beam from the 12 GeV upgraded CEBAF under construction now. This will require construction
of a hadron/nuclear beam facility to be built next to the upgraded CEBAF complex to enable such collisions.
\end{enumerate}

With the experimental conditions available at the EIC:  a) center of mass energy ($\sim 100-140$ GeV),  
b) $\sim 100-1000$ times  larger luminosity in e-p collisions than HERA,  c) polarization in electron and 
proton/deuteron/helium beams, and  d) a comprehensive detector system, 
it is only natural to explore what measurements would be possible at the EIC in the electroweak physics sector 
and of possible physics beyond the Standard Model (BSM). Three possible physics topics were considered so far: 

\begin{enumerate}
\item Precision measurement $\sin^{2}{\theta}_{W}$ as a function of Q, the momentum transfer, in e-p collisions \cite{sin2theta}. 
This would be the next generation experiment beyond the SLAC-E158, 6 GeV Q-Weak, 6 GeV PVDIS \& the currently planned 
12 GeV experiments (Moller \cite{moller} \& SoLID-PVDIS \cite{solid}) and would be complementary to atomic parity 
violation searches planned in the future.  The Q range of the EIC would be between the fixed target experiments and
the measurements from LEP at the Z-pole. Any deviation from the expected running of the $\sin^{2}{\theta}_{W}$ would be 
a hint of physics beyond the SM.

\item Possible exploration of charged lepton flavor violation, particularly transitions between the 1$^{st}$ and the 
3$^{rd}$ generation leptons ($e-\tau$) \cite{etau} at the highest energy and highest luminosity e-p collisions \cite{lpq} at the EIC. 
This would extend searches made at HERA, in different ranges of the lepto-quark couplings and masses, and will be 
complementary to future searches at the LHC, LHeC and the Super-B factories. The $\sim$100-1000 times more 
luminous collisions compared to HERA will be key to success in these searches. The angular correlations in the final
state decay particles in the known Standard Model interactions involving $\tau$'s in the final state and those in which
$\tau$s are created in a leptoquark interactions will form the tell-tale signs of the existence of leptoquarks interactions. 
Such studies were performed at HERA in the last decade, as such the requirements of detector acceptance and the
methods of analyses are fairly well defined\cite{lpqhera}.  

\item Exploration of nucleon spin structure using the electroweak probes i.e. Z (and its interference with $\gamma$ ) 
and W$^{+/-}$\cite{ewsf}. Due to their different couplings to the quarks and anti-quarks, these measurements will enable 
us to explore different combinations of partons and hence allow a determination of parton distribution functions different 
from those which are accessible through conventional deep inelastic scattering with virtual photons.  There is ample
experience with W production in e-p scattering from HERA, where detailed studies of the structure function $xF_{3}$ have
been performed. Since the HERA proton beam was unpolarized, only unpolarized heavy quark distributions were ever
measured. With the polarized proton and neutron (via either deuteron or helium) beams at the EIC, these studies can
be extended to include not only the quark anti-quark polarization but one could also study in detail their possible charge 
symmetry relations.
\end{enumerate}

The above studies and considerations are preliminary. Detailed detector simulation are needed to confirm feasibility of 
these, and are underway. 

%There might other topics of interest to the high-energy physics community possible with a machine
%like EIC. We welcome your input and thoughts on this.

%%%%%%%%%% Other Symmetry Tests %%%%%%%%%%%%%%%%%%%%

\section{Other Symmetry Tests}
\label{sec:Other}

Numerous tests of fundamental symmetries nucleons, nuclei, and atoms  -- out side the main thrusts described above -- are being pursued by members of the community. Among those for which descriptions were provided include studies of anti-matter and possible novel tests of time-reversal invariance. 

%%%%%%%%%%%% Antihydrogen ALPHA %%%%%%%%%%%%%%%%%%%

\subsection{Antihydrogen: ALPHA}
\label{sec:alpha}

The goal of the Antihydrogen Laser Physics Apparatus (ALPHA) collaboration is
to conduct fundamental studies of matter-antimatter asymmetry. 
Until our recent successes~\cite{1,2} with the current ALPHA apparatus, no group had
ever trapped neutral antimatter. In 2009, the collaboration first had a hint of success,
reporting~\cite{3} 6 'candidate' antihydrogen atoms.  In 2010, they 
demonstrated trapped antihydrogen, reporting  38~\cite{1} antiatoms with trapping times of  0.17 s, 
subsequently optimizing performance and increasing tjhe trapping numbers to  309~\cite{2} trapped antiatoms during the 2010 antiproton
season, and  hundreds more during this season, while simultaneously extending the
confinement time of these antiatoms from 0.17 s to 1000 s. These results received significant coverage in the scientific press and were enthusiastically received by the general public~\cite{3a}. The ALPHA apparatus was primarily designed to
demonstrate that cold antihydrogen could be trapped, and this has been accomplished.  Motivated by this rapid
progress towards achieving conditions required for physics studies of antihydrogen, the collaboration is rebuilding our experiment. The new apparatus, 
ALPHA-II,  is  specifically designed to be a production machine on which physics studies of antimatter are conducted. 

%The 1000 s lifetime trapped antiatom, in particular, is very encouraging. It is more than
%enough time for any of our planned physics measurements, and, critically, it
%far exceeds the time necessary for the antiatoms to decay to their ground
%state. None of the hundreds of millions of untrapped antihydrogen atoms  previously
 % created by the ATHENA~\cite{4a}, ATRAP~\cite{4b}, ALPHA~\cite{1,2}, and ASACUSA~\cite{4c}  experiments likely
%lived long enough to decay to the ground state; they were in
%unknown,  highly excited atomic states. It is very difficult to perform
%either laser or microwave spectroscopy with atoms in unknown excited states.
% Furthermore, antiatoms in excited states experience Stark (polarization) forces,   making it very difficult to
%conduct gravitational measurements.

%We took many substantial technical risks building the original ALPHA apparatus. Demonstrating trapping  validated many of these  technical design choices,
%such as using an octupole-mirror magnetic field to create the antihydrogen
%trapping potential, and conducting trapping tests via a fast turn-off of these
%fields. Trapping required the integration of ideas and techniques from beam,
%plasma, detector, and atomic physics. These include new diagnostics for low
%temperature nonneutral plasmas, evaporative cooling, and autoresonant mixing
%of antiprotons and electrons. 

%While the  success received much scientific recognition~\cite{3}, 
The success to date is only a
first step on a long road of scientific exploration. Lessons from the five years
of operation of the ALPHA apparatus are  being incorporated into the new apparatus.
ALPHA-II will be  designed, constructed, and commissioned
during the next three years. The  ALPHA-II design
will allow one to perform  microwave and laser spectroscopy,
antihydrogen charge neutrality measurements, and antimatter gravity studies. Though unlikely,  any
discovery of a fundamental difference between matter and antimatter in CPT or
gravity experiments may shed light on baryogenesis or dark energy.

Key to these future physics studies on ALPHA-II is improving its performance over that of ALPHA. The ALPHA-II design goals are  (a) a significant increase in trapping
rates, from a peak rate of about 4 antihydrogen atoms per hour to a peak of order 40
antihydrogen atoms per hour,   (b) improved trap geometry allowing for enhanced
diagnostics and laser and microwave spectroscopic access,  and (c)  improved shot-to-shot reliability.
ALPHA-II will  incorporate  the techniques that we have developed on the original ALPHA apparatus,
including  evaporative cooling~\cite{evaporative} of antiprotons and positrons, autoresonant
injection~\cite{AR} of antiprotons into the positron plasma, rotating wall plasma compression~\cite{compression} and an octupole-mirror magnetic-well~\cite{Andresen2010141} neutral confinement scheme.
The experiment will increase the production rate by separating catching and mixing functionalities,  by enhanced antiproton capture, by creating and maintaining lower positron, antiproton, and electron
plasmas, and by developing new simulation tools for fast optimization of a wide range of system parameters. The collaboration is  working on improved  plasma diagnostics to reliably measure temperature at sub-Kelvin resolution, and on new, preferably non-invasive, temperature diagnostics. 

The AD at CERN is the only facility that can provide low energy antiprotons for trapping experiments. The AD currently provides $\sim$5 MeV antiprotons to the experiments. ALPHA sends them through a degrader and traps about one in $10^3$. CERN is building a new decelerator ring which will take the AD beam and lower its energy, while providing the needed additional cooling, to 100kV. The construction of ELENA should begin in 2013 and the first physics injection should follow about three years later.  This will significantly enhance antiproton capture rates, and, consequently,  antihydrogen production. New  positron-antiproton mixing schemes may be developed to best utilize increased antiproton numbers. 

 Antimatter experiments have one advantage over experiments with normal matter: One can detect, with over 60\% efficiency, the interaction of each antihydrogen on our trap wall. Thus, with only a few antiatoms, one can perform experiments that would be impossible with normal atoms.  For example, a microwave-induced spin flip  to a high field
seeking state can be detected on almost a per atom basis. If the flip is successful, the antihydrogen hits the trap wall and the resulting shower is detected. The detector has both spatial and temporal resolution and serves as a critical diagnostic for inferring the temperature of the trapped antiatoms.

\subsection{Antiprotons and Antihydrogen}
\label{sec:antip}

\noindent{\bf Matter-antimatter symmetry and CPT}

The discrepancy between the deviation from matter-antimatter symmetry on the cosmic scale on one hand and the so far observed perfect symmetry between particle and antiparticle properties on a microscopic scale on the other hand is one of the big mysteries that has not yet been satisfactorily explained by the standard model (SM) of particle physics. The observed baryon asymmetry in the cosmos of $(N_{B}-N_{\overline{B}})/N_{\gamma} \sim 10 \times10^{-6}$ \cite{Dolgov:2009fk} is in the SM thought to originate from the three Sakharov conditions 
{\em i)} Baryon number violation, 
{\em ii)} C and CP symmetry violation, and
{\em iii)} deviations from thermal equilibrium during the expansion of the universe.
The so-far observed violations of CP symmetry in the $K$ and $B$ meson sector are however too small to quantitatively explain the observed baryon asymmetry and thus other sources of matter-antimatter asymmetry may be explored. Notably, if CPT symmetry is violated, then Sakharov's third criteria need not be invoked. As a concrete illustration, one may consider the Standard Model extension by Kostelecky \emph{et al.} \cite{Colladay:1997vn}, wherein it is possible to generate a large baryon asymmetry through violations of CPT \cite{Bertolami:1997uq}.

 CPT symmetry ensures that particles and antiparticles have perfectly equal properties. It is the result of a proof based on mathematical properties of the quantum field theories used in the SM, but certain of these properties like point-like particles are not any more valid in extension of the SM like string theory. Thus tests of CPT by precisely comparing particle and antiparticle properties constitute a sensitive test of physics beyond the SM. Antiprotonic atoms and especially antihydrogen offer the most sensitive tests of CPT in the baryon sector.

%The gravity of antimatter

\noindent{\bf CPT tests with antiprotonic atoms and antihydrogen}

Since more than 20 years, low-energy antiprotons have provided the most sensitive tests of CPT  in the baryon sector. The TRAP collaboration at LEAR of CERN has determined the maximal deviation of the charge-to mass ratio of proton and antiproton to 
$(Q_{\overline{p}}/M_{\overline{p}})/(Q_{p}/M_{p})+1 = 1.6(9)\times 10^{-10}$
 \cite{Gabrielse:99,Thompson:2004fk} using a Penning trap. The ASACUSA collaboration at CERN's Antiproton Decelrator has been performing precision laser and microwave spectroscopy of antiprotonic helium, an exotic three-body system containing a helium nucleus, an antiproton and an electron exhibiting highly-excited metastable states \cite{Hayano:2007}. By comparing the experimentally observed transitions frequencies between energy levels of the antiproton with state-of-the art three-body QED calculations the most precise values for the equality of proton and antiproton mass and charge ( $  (Q_{p}+Q_{\overline{p}})/Q_{p} = (M_{p}-M_{\overline{p}})/M_{p}< 7\times10^{-10}$ \cite{Hori:2011gk}) and the antiproton magnetic moment (($\mu_{p}-\mu_{\overline{p}})/\mu_{p} < 2.9\times10^{-3}$ \cite{Pask:2009lq}) have been obtained.
 
Antihydrogen ($\overline{\mathrm{H}}\equiv \overline{p}e^{+}$), the simplest atom consisting only of antimatter, promises the highest sensitivity because its CPT conjugate system, hydrogen, is one of the best studied atoms in physics. Currently three collaborations at CERN aim at forming antihydorgen and perform precision spectroscopy of its structure: ATRAP and ALPHA have the goal to measure the two-photon 1S--2S laser transition in antihydrogen, that has been determined to a relative precision of $10^{-14}$ in hydrogen, and ASACUSA aims at measuring the 1.4 GHz ground-state hyperfine transition that is known to $10^{-12}$ relative precision from the hydrogen maser. 

The progress -- as is typical for precision experiments -- is slow: the first formation of antihydrogen was reported in 2002 by ATHENA \cite{ATHENA-Hbar:02} (the predecessor of ALPHA)  and ATRAP \cite{ATRAP-Hbar:02a} using a nested Penning trap technique \cite{Gabrielse:2005sl}, the next major step happened in 2010 when ALPHA reported the first trapping of neutral antihydrogen atoms in a Joffe-Pritchard trap  \cite{Andresen:2010jba} (and later announced trapping times of 1000 seconds 
\cite{Andresen:2011jg}), and ASACUSA announced the first formation of antihydrogen in a different trap named ``cusp trap'' \cite{Enomoto:2010uq} which is expected to provide a polarized $\overline{\mathrm{H}}$ beam \cite{Mohri:2003wu} useful for measuring the ground-state hyperfine structure in an atomic beam \cite{Juhasz:2006uq}. Given this major achievements, first spectroscopy results can be expected within the next few years. 

To reach the full potential of the measurements, \emph{i. e.} a similar precision than was obtained in hydrogen, a much longer time will be needed. The currently only facility in the world providing low-energy antiproton, the AD at CERN, will be upgraded by an additional storage ring ELENA to decelerate antiprotons further and thus increase the number of trapped antiprotons. By the end of this decade another facility called FLAIR might get into operation at the FAIR facility under construction at Darmstadt.

\noindent{\bf Gravity of antimatter}

Using neutral antihydrogen, the gravitation between matter and antimatter can be experimentally investigated for the first time. Scenarios exist where a difference in the gravitational interaction of matter and antimatter can arise, which are part of a general search for non-Newtonian gravity \cite{Nieto:1991uq}. The AEgIS experiment has been approved at CERN-AD and is about to start just now, aiming initially at a \%-level measurement of the garviational acceleration of the antiproton. A second collaboration GBAR has submitted a letter of intent and is preparing a proposal, so that enhanced activities in this field can be expected in the future.

\def\Discussion{\setlength{\parskip}{0.3cm}\setlength{\parindent}{0.0cm}
     \bigskip\bigskip      {\Large {\bf Discussion}} \bigskip}\def\speaker#1{{\bf #1:}\ }
\def\endDiscussion{} 
\end{document}